\documentclass{elsart}
\usepackage{graphicx}
\usepackage{amsmath}
\usepackage[usenames]{color}
\usepackage{natbib}
\usepackage{lineno}

\def\deg{{^\circ}}
\newcommand{\Ro}{\text{Ro}}
\newcommand{\E}{\text{E}}
\newcommand{\Rm}{\text{Rm}}

\newcommand{\Ha}{\text{Ha}}
\renewcommand{\Re}{\text{Re}}
\newcommand{\N}{\text{N}}

\begin{document}

\begin{frontmatter}

\title{Rapidly rotating spherical Couette flow in a dipolar magnetic
  field:
  an experimental study of the mean axisymmetric flow}

\author{Nataf, H.-C.\corauthref{Nataf}}
\ead{Henri-Claude.Nataf@ujf-grenoble.fr}
\ead[url]{http://www-lgit.obs.ujf-grenoble.fr/recherche/geodynamo/Epage.html}
\author{Alboussi\`ere, T.,}
\author{Brito, D.,}
\author{Cardin, P.,}
\author{Gagni\`ere, N.,}
\author{Jault, D.,}
\author{and Schmitt, D.}

\corauth[Nataf]{corresponding author}

\address{Geodynamo team, LGIT-UMR5559-CNRS-UJF, Grenoble, France}


\begin{abstract}
In order to explore the magnetostrophic regime expected for planetary cores, in which the Lorentz forces balance the Coriolis forces, experiments have been conducted
in a rotating sphere filled with liquid sodium, with an imposed dipolar magnetic field (the $DTS$ setup). The field is produced
by a permanent magnet enclosed in an inner sphere, which can rotate at a separate rate, producing a spherical
Couette flow. The flow properties are investigated by measuring electric potentials on the outer sphere, the
induced magnetic field in the laboratory frame just above the rotating outer sphere, and velocity profiles inside the liquid sodium using
ultrasonic Doppler velocimetry. The present article focuses on the time--averaged axisymmetric part of the flow. The
electric potential differences measured at several latitudes can be linked to azimuthal velocities, and
are indeed found to be proportional to the azimuthal velocities measured by Doppler velocimetry. The Doppler profiles
show that the angular velocity of the fluid is relatively uniform in most of the fluid shell, but rises near
the inner sphere, revealing the presence of a ``magnetic wind'', and gently drops towards the outer sphere.
The transition from a magnetostrophic flow near the inner sphere to a geostrophic flow near the outer sphere is controlled by
the local Elsasser number. For Rossby numbers up to order 1, the observed velocity profiles all show a similar shape.
Numerical simulations in the linear regime are computed, and synthetic velocity profiles are compared with the measured ones. A good agreement is found for the angular velocity profiles. In the geostrophic region, a torque--balance model provides very
good predictions. Radial velocities
change sign with the Rossby number, as expected for an Ekman--pumping dominated flow.
For a given Rossby number the amplitude of the measured angular velocity is found to vary by as much as a factor of 3.
Comparison with numerical simulations suggests that this is due to variations in the electric coupling between liquid sodium
and the inner copper sphere, implying an effect equivalent to a reduction of the inner sphere electric conductivity by as much as a factor 100.
We show that the
measured electric potential difference can be used as a proxy of the actual fluid velocity. Using this
proxy in place of the imposed differential velocity, we find that the induced magnetic field varies
in a consistent fashion, and displays a peculiar peak in the counter--rotating regime. This happens when
the fluid rotation rate is almost equal and opposite to the outer sphere rotation rate. The fluid is then
almost at rest in the laboratory frame, and the Proudman--Taylor constraint vanishes, enabling a strong meridional
flow. We suggest that dynamo action might be favored in such a situation.
\end{abstract}


\end{frontmatter}


\section{Introduction}
\label{Introduction}

The internal dynamics of liquid planetary cores is deeply influenced by the rotation of the planet and (in many cases) by the presence of a
magnetic field, and can be in the so--called magnetostrophic regime where the Coriolis and Lorentz forces balance each other. In recent years, attention has been paid to the rotating spherical Couette flow in the presence
of a magnetic field \citep{Kleeorin97,Hollerbach94,Dormy98,Starchenko98,Hollerbach00,Nataf06}. An electrically conducting liquid fills the gap
between two concentric spheres and is sheared through the differential rotation of the two spheres. A magnetic
field can be applied.
This flow has the distinct advantage of being amenable
to both experiments and numerical models, while retaining important ingredients of natural situations. 
Numerical simulations \citep{Dormy98,Hollerbach00} and asymptotics \citep{Kleeorin97,Starchenko98,Dormy02} highlight the importance of the magnetic wind that arises when the magnetic field
lines are not parallel to the rotation axis, as for a dipolar field. This magnetic wind can entrain the fluid
at angular velocities larger than that of the inner sphere, a phenomenon called super--rotation \citep{Dormy98,Dormy02}.
The rotating
spherical Couette flow has also been proposed for producing a dynamo \citep{Cardin02}. Numerical models have shown
that dynamo action could be produced this way, even in liquids with a low magnetic Prandtl number, such as liquid metals 
\citep{Schaeffer06}. This
has led Dan Lathrop and his team to build a 3m--diameter sphere with a rotating inner sphere in the hope of
starting a dynamo, once the gap is filled with liquid sodium and the differential rotation is large enough.

Here, we report on results obtained in the $DTS$ ({\it Derviche Tourneur Sodium}) experiment. This experiment, described in \cite{Cardin02} and \cite{Nataf06},
is a rotating spherical Couette flow with an imposed axisymmetric dipolar magnetic field. Because of this strong imposed field, and although
liquid sodium is used, the dynamo effect cannot be studied in the $DTS$ experiment. However, it permits to explore very similar
dynamics. In particular, the magnetic field is large enough to strongly influence the flow (interaction
parameter $\N$ up to 200) and the magnetic field is strongly affected by the flow (magnetic Reynolds number $\Rm$ up to 44).

This kind of experiment can bring information on the level and organization of turbulence in situations where
both rotation and the magnetic field impose strong constraints on the flow. Local theories have been proposed \citep{Braginsky90}, partly backed by numerical simulations \citep{StPierre96}, and some teams are trying to implement
sub-grid models based on these theories in full dynamo models \citep{Buffett03,Matsui07}. Results obtained in the $DTS$ experiment
suggest an alternative view, in which waves play a major role \citep{Schmitt08}. A good characterization of
the mean axisymmetric flow is needed in order to fully exploit these observations.

An experimental setup, built by Dan Lathrop and his group \citep{Sisan04,Kelley07} is very similar to our $DTS$ experiment.
The main
difference is that an axial magnetic field is applied by external coils in \cite{Sisan04} and \cite{Kelley07}, while we impose a strong dipolar magnetic field in $DTS$ since the inner sphere contains a permanent
magnet. One of the consequences is that the entrainment of the conducting fluid is much stronger, since the inner sphere
acts as a magnetic stirrer.
We also have developed specific instrumentation, such as measuring the electric potential at the surface
of the outer sphere, which carries important information on the flow \citep{Nataf06,Schmitt08}.


In this article, we focus on the time--averaged axisymmetric magnetohydrodynamic flow observed in $DTS$ when the outer sphere
rotates. The first observations were described in \cite{Nataf06}, and more thorough results for a static outer sphere
will be discussed in a forthcoming article. 

\cite{Nataf06} showed that the measurement of electric potential differences at the surface
of the outer sphere could be related to the azimuthal velocity of the liquid beyond the Hartmann boundary layer. It was
found that this velocity could be larger than the solid--body rotation of the inner sphere, providing experimental
evidence for the so--called super--rotation
predicted by \cite{Dormy98}. However, the latitudinal variation of the electric potential differences was found to differ
markedly from the one calculated in a linear numerical model, analogous to that of \cite{Dormy98}. The relative variation
of electric potential differences with latitude turned out to be universal, independent of the differential rotation
of the inner sphere. This was particularly true when the outer sphere was at rest. The electric design of the electrodes
and acquisition system has been improved since, yielding better quality signals.

\begin{figure}
\centerline{
\begin{tabular}{ccc}
\includegraphics[width=7.3cm]{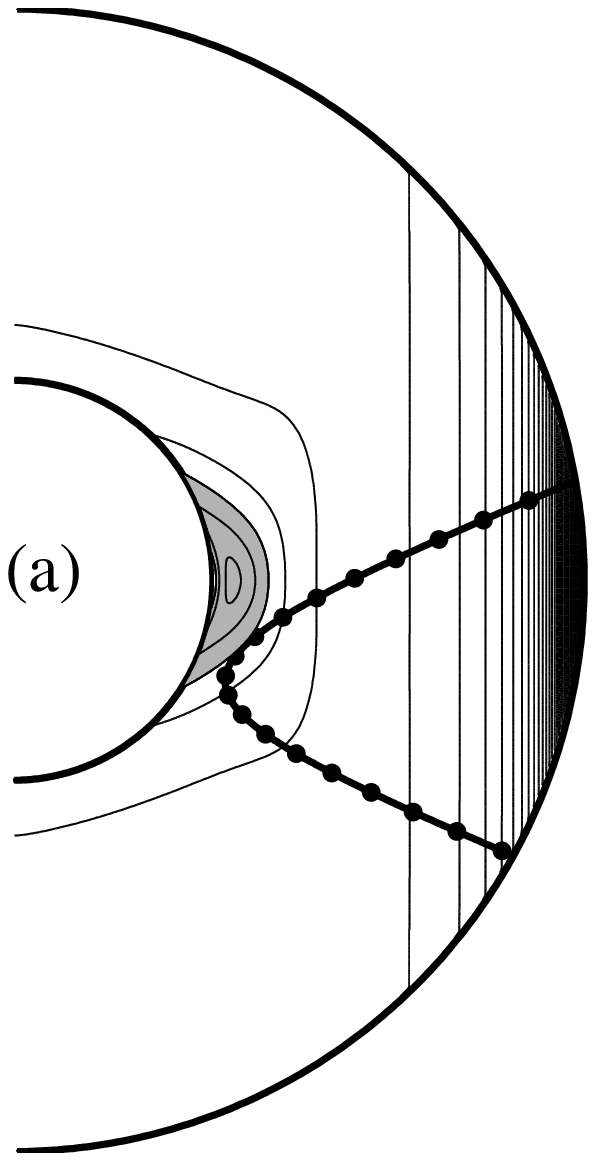} & \hspace{1cm} &
\begin{minipage}{0.3\linewidth}
\vspace{-12cm}
\includegraphics[width=6cm]{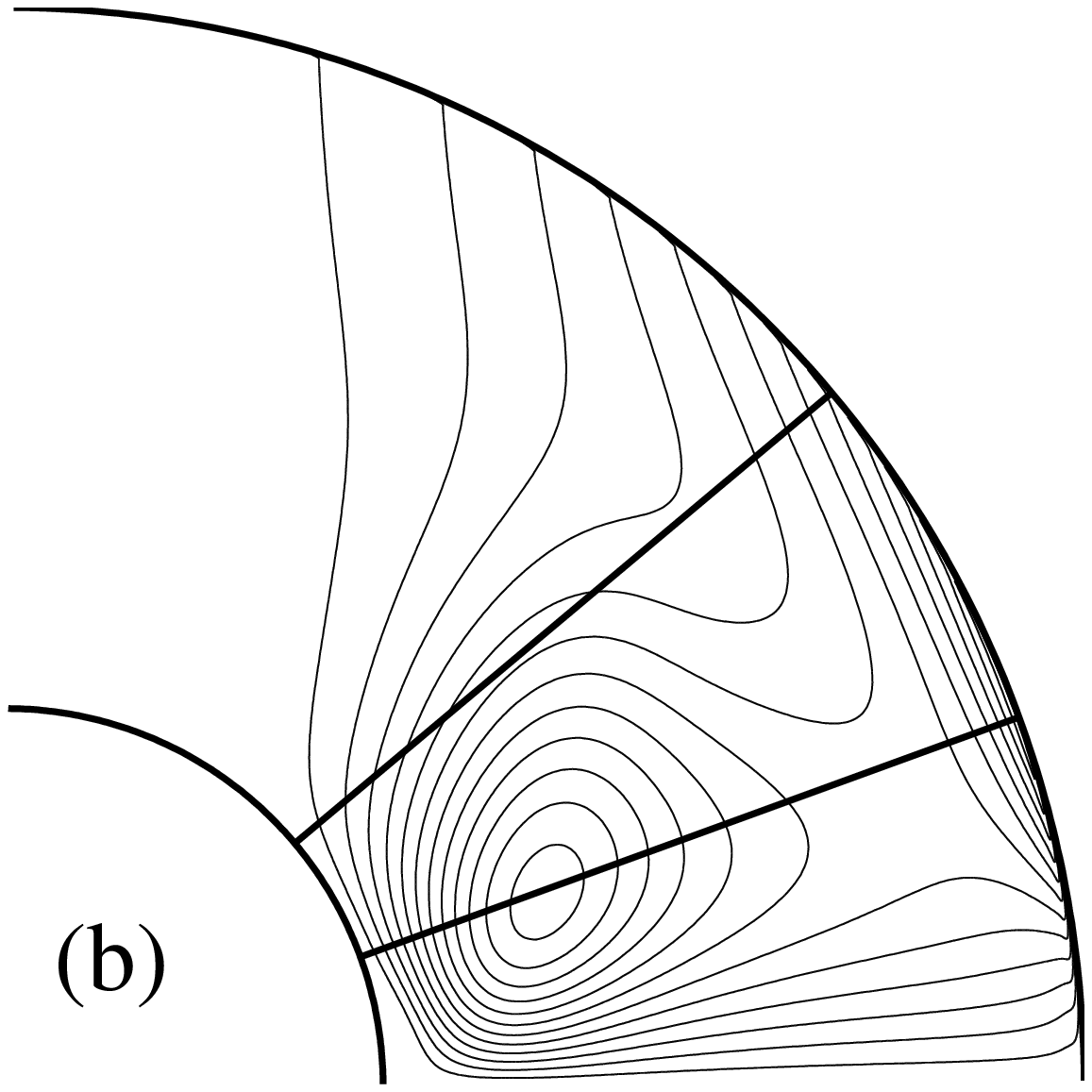}
\includegraphics[width=6cm]{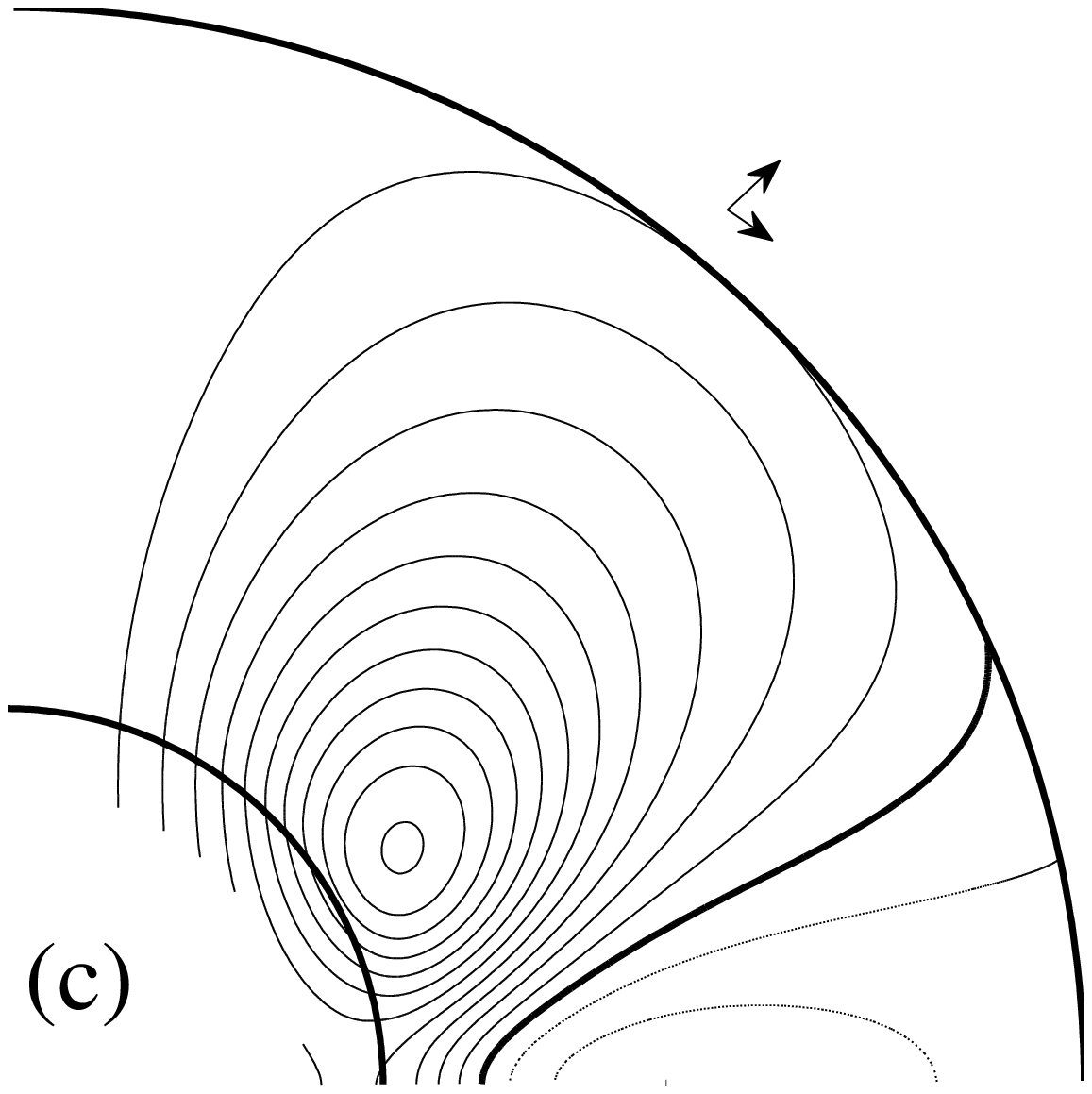}
\end{minipage}
\end{tabular}
}
%
%
     \caption{Non--linear axisymmetric numerical simulation with the geometry and parameters of the $DTS$ setup, corresponding to $f =5$ Hz
and $\Delta f=0.4$ Hz (see section \ref{Numbers} below) (a) Contour map of the angular velocity. The interval between contours is 0.027 in $2 \pi \Delta f$ units. Angular velocities are larger than that of the inner sphere in the shaded region (super--rotation).
Two different regions are clearly visible: the region near the inner sphere, where the flow lines are
aligned with the magnetic field lines of the imposed dipole (Ferraro law), and where the angular velocity exceeds that of the inner sphere by up to $6 \%$; farther away, a region where the flow is nearly geostrophic, with contours aligned
with the rotation axis (vertical). Also drawn, the trajectory of the ultrasonic beam shot from the ``azimuthal'' assembly at $+10\deg$, projected on the meridional plane. Marks are drawn every 2 cm along the ray path. (b) Meridional
circulation. The interval between contours is $2.7 \times 10^{-5}$ in $2 \pi a^2 \Delta f$ units, where $a$ is the radius
of the outer sphere. The main circulation is counterclockwise, {\it i.e.} centrifugal at the equator. It changes sign with $\Delta f$. Radial velocities are measured along radial shots at latitudes $40\deg$ and $20\deg$ (lines). (c) Field lines of the induced meridional magnetic field. The interval between contours is $2 \times 10^{-3}$ in
$a B_0$ units. $B_0$ is the intensity of the imposed dipolar magnetic field at the equator at the outer sphere. The 0 line is thicker. While the imposed dipole is pointing downwards, the induced field points up at high latitudes. $B_r$ and $B_{\theta}$
are measured at a latitude of $50 \deg$, 1 cm away from the sodium boundary (arrows).}
        \label{fig:numerics}
\end{figure}

Numerical models provide a useful guide for interpreting the experimental observations. Linear solutions
have been computed by \cite{Dormy98} for a situation similar to ours (spherical Couette with an imposed
dipolar magnetic field, a conducting inner sphere and a rapidly rotating outer sphere). The results show that as the magnetic field is increased, the flow evolves from the classical Stewartson solution \citep{Stewartson66},
with a cylindrical shear--layer attached to the rotating inner sphere, to a solution where the fluid
enclosed within a closed magnetic field line is entrained by the inner sphere to angular velocities in excess
of the imposed differential rotation. A shear--layer develops around that field line \citep{Dormy02}.
Using the numerical approach described in \cite{Nataf06}, we have extended these computations to conditions
closer to the actual $DTS$ configuration, with stronger global ($f$) and differential ($\Delta f$) rotation rates. The results
are presented in figure \ref{fig:numerics}. The angular velocity contours (figure \ref{fig:numerics}a) clearly illustrate the existence
of two domains: a region close to the inner sphere where the magnetic effects dominate, and a region where
the flow is geostrophic farther away. In the inner region, the angular velocity contours follow the
magnetic field lines (Ferraro law of isorotation \citep{Ferraro37}), while they are aligned with the rotation axis in the geostrophic region.
\cite{Dormy98} demonstrate that the radial extent of the magnetic--dominated
region solely depends upon the Elsasser number (which compares the Lorentz and Coriolis forces). In contrast
to the Stewartson solution, the bulk of the fluid is entrained by the inner sphere, and a thin Ekman--Hartmann
layer forms beneath the outer shell. Ekman pumping results and feeds the meridional circulation shown
in figure \ref{fig:numerics}b. The circulation around the main vortex is counterclockwise in the upper hemisphere
({\it i.e.} centrifugal at
the equator) when the inner and outer spheres rotate in the same direction, and clockwise ({\it i.e.} centripetal
at the equator) when the two spheres are in contra--rotation. The field lines of the meridional magnetic field produced by this
circulation are drawn in figure \ref{fig:numerics}c.


Direct measurements of the flow velocity were presented in \cite{Nataf06}, using ultrasonic Doppler velocimetry,
but only radial profiles of the radial velocity could be obtained. More profiles are now available and can be compared
with the numerical simulations. The setup now permits the record of profiles of the angular
velocity as well, revealing the actual variation of angular velocity inside the sphere. We will see that they confirm
the progressive rise of the angular velocity from the equator of the outer sphere to the interior of the fluid predicted
by the numerical simulations (figure \ref{fig:numerics}a). The existence of this peculiar boundary layer was predicted by
\cite{Kleeorin97}. We will present their approach and propose an extension that provides a better fit to our observations.

There were some hints in \cite{Nataf06} that different flows could be observed for the same imposed velocities. With additional
data now available, this tendency is fully confirmed. We think that this is due to variations in the electric coupling
between the liquid sodium and the copper inner sphere. 
Indeed, the entrainment of the liquid is not viscous but depends on electric currents that flow between the liquid sodium and the copper shell of
the inner sphere. The presence of oxides or other impurities, or imperfect wetting on that shell can hinder the electric currents.
However, we show in this article that we can use the difference in electric potentials at one latitude ($40 \deg$ for example) as a proxy of the
actual differential angular velocity of the fluid, and get a much more coherent picture of the various observations, if we
relate them to this actual value.

No measurement of the induced magnetic field was provided by \cite{Nataf06}. Here, we present measurements of the radial and orthoradial components of the induced magnetic
field at a latitude of about $50 \deg$ just outside the outer sphere, in the laboratory frame. The mean values correspond
to the field induced by the mean meridional flow. When the outer sphere is rotating, a very peculiar behavior is monitored
for a particular counter-rotation rate of the inner sphere.

We recall the main features of the $DTS$ setup and measuring techniques in section \ref{Setup}. The typical dimensionless numbers
are discussed in section \ref{Numbers}. The results are presented in section \ref{Results}, and discussed in section \ref{Discussion}.

\section{Experimental setup and measuring techniques}
\label{Setup}

\subsection{Setup}

\begin{figure}
  \centerline{ \includegraphics[width=10cm]{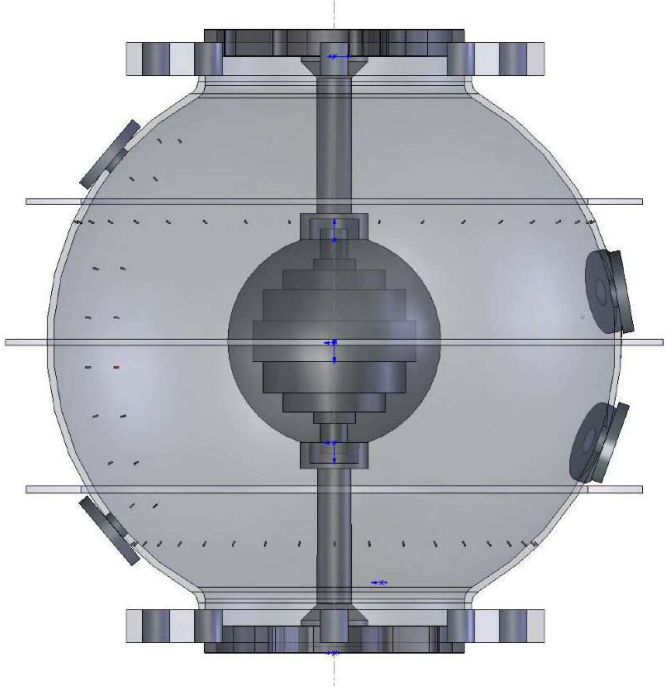}}
     \caption{Sketch of the central part of the $DTS$ experiment. The 7.4 cm-radius inner sphere is made of copper and contains 5 layers of magnets.  The inner radius of the stainless steel outer sphere is 21 cm and its thickness is 5 mm. Blind holes are drilled and threaded along several meridians and parallels to receive M1.2 mm brass bolts acting as electrodes. In this article, results are presented for electric potentials measured along one meridian (on the left). The large holes at latitudes $-20 \deg$, $10 \deg$ and $\pm 40 \deg$ receive interchangeable assemblies, which can be equipped with ultrasonic transducers. Magnetometers are placed
at a latitude of $50\deg$, just above the outer sphere, in the laboratory frame. They measure the radial $B_r$ and orthoradial 
$B_{\theta}$ components of the magnetic field.}
        \label{fig:sphere}
\end{figure}

The central part of the $DTS$ experiment is sketched in figure \ref{fig:sphere}. Forty liters of liquid sodium are contained between an inner sphere of radius $b$ = 7.4 cm 
and an outer spherical surface of radius $a$ = 21 cm. The 5 mm--thick outer sphere is made of stainless steel, which is 8 times less electrically conductive than sodium at $130\deg$C, the
typical experimental temperature. The inner sphere is made of two hollow copper hemispheres filled with 5 layers of magnetized rare-earth cobalt bricks. Copper is about 4.2 times more conductive than sodium. Outside the copper
sphere, the magnetic field is found to be an axial dipole, aligned with the axis of rotation and pointing downwards with:
$$
\vec{B}(r,\theta) = \frac{\mu_0 {\cal M}}{4 \pi r^3}(2 \vec{e}_r \cos\theta + \vec{e}_{\theta} \sin\theta),
$$
in spherical coordinates, with $\theta$ the colatitude and ${\cal M} = -700$ Am$^2$ the magnetic dipolar moment, yielding a magnetic field amplitude ranging from 0.345 T at the poles
of the inner sphere down to 0.008 T at the equator of the outer sphere. More details are given in \cite{Nataf06}.

The inner and outer spheres are set in rotation around the vertical axis. In this article, we present results obtained for imposed rotation rates $f$ of the outer sphere of
5, 10 and 15 Hz. The rotation rate of the inner sphere in the laboratory frame is varied between -25 and +25 Hz, yielding differential rotation rates $\Delta f$ of the inner
sphere in the rotating frame ranging from -40 to 20 Hz (depending upon $f$). Each sphere is entrained by an 11 kW brushless electric motor (Yaskawa SGMGH-1ADCA61).

The torques and actual rotation rates are retrieved as analog signals from the two motor controllers. In order to decipher the organisation of the flow in the liquid,
several techniques are employed. Electric potentials are measured at several locations on the outer sphere. Profiles of the flow velocity are obtained using ultrasonic
Doppler velocimetry. Since the outer sphere is rotating, all these signals must pass through electric slip-rings to be analyzed in the
laboratory frame. We use a 36--tracks 10A slip--ring system (Litton AC6275--12). In order to decrease electromagnetic perturbations of the signals, this system is completely enclosed
in an aluminum casing. In contrast, the magnetic field is measured in the laboratory frame, with magnetometer sensors a few millimeters away from the rotating surface of the outer sphere.

\subsection{Electric potentials}

We measure the electric potential at several points on the outer sphere. Blind
holes are drilled in the 5 mm--thick stainless steel shell. The holes are 1 mm
in diameter, and 4 mm--deep. They are threaded and equipped with 3 mm--long M1.2 mm brass
bolts, on the head of which electrodes are soldered. Stainless steel is a poor
electric conductor as compared to liquid sodium and thus it does not affect
the dynamically generated electric potential, but it is a good enough electric
conductor for the impedance of electrode pairs to be much smaller than that of
the acquisition system. The measurements are acquired at 1 kHz with a PXI-6229 National Instruments board,
after going through a simple anti--aliasing 215 Hz low--pass RC filter. As we focus here on the
mean axisymmetric state, these data are time--averaged over time--windows ranging from 0.1 to 1 second, in
order to remove fluctuations.

In this article, we present data for the difference in electric potential
measured between electrodes placed along the same meridian. These electrodes
are $10\deg$ apart and range from $-45\deg$ to $+45\deg$
in latitude as shown in figure \ref{fig:sphere}. Assuming steady state, the difference in
electric potential $\Delta V$ between two electrodes separated by a
latitudinal angle $\Delta \theta$ is related to the azimuthal velocity
$U_{\varphi}$ of the liquid sodium by:
\begin{equation}
  \frac{\Delta V}{a \Delta \theta} = U_{\varphi} B_r -
  \frac{j_{\theta}}{\sigma},
  \label{eq:ohm}
\end{equation}
where $B_r$ is the radial component of the magnetic field, $j_{\theta}$ is
the $\theta$-component of the electric current, and $\sigma$ is the electric conductivity of liquid sodium.

In \cite{Nataf06}, we used the differences in electric potential to deduce the angular velocity of the liquid
sodium flow beneath the Hartmann boundary layer. Indeed, assuming that the meridional electric currents vanish
below that layer, equation \ref{eq:ohm} yields:

\begin{equation}
U_{\varphi} = \frac{\Delta V}{a \Delta \theta \, B_r} \; \; \text{or} \; \; \Delta f_\text{fluid} (s) = \frac{\Delta V}{2 \pi s \, a \Delta \theta \, B_r},
\label{eq:uphi}
\end{equation}

where $\Delta f_\text{fluid} (s)$ is the angular frequency of the fluid with respect to the outer sphere at
the cylindrical radius $s$ for the latitude of the electrode pair.
In this article, we will use the difference in electric potential
measured at $40\deg$ latitude ($\Delta V_{40\deg}$) as a proxy of the actual rotation frequency of the
liquid sodium flow with respect to the outer sphere.

\subsection{Ultrasonic Doppler velocimetry}

Since liquid sodium is opaque, it is not possible to use optical methods to
investigate the flow within the sphere. Instead, we use ultrasounds to probe
the flow. Ultrasonic transducers (TR0405HS from Signal Processing, Switzerland) are placed in removable assemblies on the
outer sphere, and shoot a narrow beam of 4 MHz ultrasounds. The pulsed Doppler
velocimetry technique is perfectly suited for retrieving profiles of the
component of the flow velocity along the shooting line, as illustrated in \cite{Brito01}, \cite{Eckert02}, and \cite{Nataf06}.
We use a DOP 2000 instrument from Signal Processing, Switzerland.

While only radial profiles of the radial velocity were discussed in \cite{Nataf06}, we show
here profiles of both the radial and angular velocities. Radial profiles are measured at two different latitudes
($-20\deg$ and $-40\deg$), as drawn in figure \ref{fig:numerics}b. The angular velocities are obtained with an assembly devised for shooting a beam at an
angle of $24\deg$ away from radial.
Radial velocities are found to be at least one order of magnitude smaller than the azimuthal velocities.
Neglecting the radial velocity component, the axisymmetric angular frequency $\Delta f_\text{fluid}(s,z)$ of the fluid flow
can thus be retrieved by projection of the
velocity ${\cal U}(d)$ measured along the shooting line. With our actual geometry, one gets:

\begin{equation}
\Delta f_\text{fluid} (s,z) = \frac{2.88}{2 \pi} \frac{{\cal U}(d)}{a},
\label{eq:umes}
\end{equation}

where $d$ is the distance along the shooting line, $s$ the cylindrical radius, and $z$ the height above the equatorial plane,
which are related by:

\begin{eqnarray}
z (d) & = & a \, \sin10\deg - 0.359 \, d \label{eq:zd} \\
s (d) & = & a \, \cos10\deg \sqrt{1 - 1.76 \frac{d}{a} + 0.898 \left( \frac{d}{a} \right)^2}
\label{eq:sd}
\end{eqnarray}

As shown in figure \ref{fig:numerics}a, plotting $z$ versus $s$ for this beam yields the projection of the ray path in a meridional
plane. Tick marks have been drawn every 2 cm along the path. The minimum cylindrical radius reached is $s = 7.7$ cm, close to the radius
of the inner sphere (7.4 cm), but it is attained some 4 cm beneath the equator. 

The Doppler technique relies upon ultrasonic energy backscattered by particles in the liquid. We did not add
any particle in the liquid sodium. In liquid metals, it is often assumed that the backscattered echoes are
due to floating oxides. It is interesting to note that we were able to record good profiles, even when the
outer sphere was rotating at 14 Hz, implying centrifugal acceleration up to 180 times the acceleration due
to gravity.

\subsection{Induced magnetic field}

The magnetic field is measured outside the sphere, in the laboratory frame. Both the radial $B_r$ and
the orthoradial $B_\theta$ components are measured with a ``Giant MagnetoResistance'' ($GMR$) chip. The two chips are mounted on a circuit
protected by a stainless steel tube. The probe has been installed at a latitude of $50\deg$, as close as possible to the rotating sphere: 
the distance to the center of the sphere is about 22 cm. The variation of temperature has a significant effect
on the measurements. Temperature is thus measured close to the $GMR$ magnetometers, and a linear correction
scheme is applied. The acquisition rate is 2000 samples per second, but all data presented here have been
moving--averaged over 3 seconds, since we are focusing on the time--averaged axisymmetric mean flow.

\section{Dimensionless numbers}
\label{Numbers}

Relevant parameters and dimensionless numbers are given in table \ref{tab:dimen}. The symbols have their usual meaning :
$\eta$ is the magnetic diffusivity, $\rho$ the density, $\nu$ the kinematic viscosity of the liquid sodium. As in all moderate--size
liquid sodium experiments, the Joule dissipation time $\tau_J$ is relatively short. Since the imposed magnetic field is dipolar,
its amplitude is much larger near the inner sphere ($B = B_i$) than near the outer sphere ($B = B_o$). We therefore evaluate those
numbers that depend on the strength of the magnetic field in these two regions. The Hartmann number $\Ha$ is large everywhere. The Lundquist number $S$ is large near the inner sphere, implying that Alfv\'en waves can play a role in the dynamics, with typical velocities $U_a$ ranging from 0.2 to 5 m/s, in the same range as flow velocities. 

Values of the Ekman number $\E$ are given for two typical values of the rotation rate $f$ of the outer sphere. The $DTS$ experiment has been designed for the study of the magnetostrophic regime, in which Lorentz forces and Coriolis forces are of
similar amplitude. This is measured by the Elsasser number $\Lambda$, which ranges from 0.01 near the outer sphere (where Coriolis forces dominate) to nearly 10 near the inner sphere (where Lorentz forces dominate).
For time--scales short compared to the magnetic diffusion time, a measure of the relative influence of the rotational and magnetic
effects is given by the dimensionless
number $\lambda$, the ratio of the Alfv\'en velocity to the inertial wave velocity \citep{Fearn88,Cardin02}, which is called the Lehnert number \citep{Lehnert54} by \cite{Jault08}. The Lehnert number is small in planetary cores, and rotational effects are expected to dominate at short time--scales. In the $DTS$ experiment, the Lehnert number is less than 1.

We estimate a typical velocity $U$ from the magnitude of the differential rotation $\Delta f $ of the inner sphere
with respect to the outer sphere. The estimate is based on the tangential velocity at the equator of the inner
sphere. We will see that, because the magnetic coupling between the magnetized inner sphere and the liquid sodium
is strong, velocities in the fluid layer typically reach this value. This yields magnetic Reynolds
numbers $\Rm$ in excess of 40, while the Reynolds number $\Re$ lies in the range $10^5-10^6$. The interaction
parameter $\N$ ranges from 0.01 near the outer sphere to 250 near the inner sphere, meaning that inertial effects
only play a role near the outer boundary. Note that the Rossby number $\Ro$ is here simply defined
as $\Ro = \Delta \Omega / \Omega = \Delta f / f$.

The $DTS$ experiment thus enables us to explore an original parameter range, which covers some aspects of liquid
core dynamics.  


\begin{table}
\begin{center}
\begin{tabular}{ccccccc}
\hline
symbol & expression & units  & \multicolumn{4}{c}{value} \\
\hline
 $a$  & outer radius & cm& \multicolumn{4}{c}{$21$} \\
 $b$ & inner radius & cm & \multicolumn{4}{c}{$7.4$} \\
 $\tau_J$ & $a^2/\pi^2\eta$ & s & \multicolumn{4}{c}{$0.05$} \\
 & & & \multicolumn{2}{c}{$\underline{\qquad B=B_i \qquad}$} & \multicolumn{2}{c}{$\underline{\qquad B=B_o \qquad}$}\\
 $\Ha$ & $aB/\sqrt{\mu_0\rho\nu\eta}$& &\multicolumn{2}{c}{4400} & \multicolumn{2}{c}{210}\\
 $S$ & $aB/\eta\sqrt{\mu_0\rho}$& &\multicolumn{2}{c}{12} & \multicolumn{2}{c}{0.56}\\
 $U_a$ & $B/\sqrt{\mu_0\rho}$&m s$^{-1}$ & \multicolumn{2}{c}{5.1} & \multicolumn{2}{c}{0.2}\\
   \hline
 & & & \multicolumn{2}{c}{$\underline{\qquad f = 5 \text{Hz}\qquad}$} & \multicolumn{2}{c}{$\underline{\qquad f = 15 \text{Hz} \qquad}$}\\
 $\E$ & $\nu/\Omega a^2$ & & \multicolumn{2}{c}{$4.7\;10^{-7}$}& \multicolumn{2}{c}{$1.6\;10^{-7}$}\\
  & & & $\underline{\: B=B_i \:}$ & $\underline{\: B=B_o \:}$&$\underline{\: B=B_i \:}$ & $\underline{\: B=B_o \:}$\\
 $\Lambda$ & $\sigma B^2/\rho \Omega$ & & 9.2 & 0.02& 3.1& 0.01\\
 $\lambda$ & $U_a/a \Omega$ & & 0.77 & 0.04& 0.26& 0.01\\
 \hline
 & & & \multicolumn{2}{c}{$\underline{\qquad \Delta f = 0.5 \text{Hz} \qquad}$} & \multicolumn{2}{c}{$\underline{\qquad \Delta f = 40 \text{Hz} \qquad}$}\\
 $U$ & $b\Delta \Omega$ &m s$^{-1}$&\multicolumn{2}{c}{0.23}& \multicolumn{2}{c}{18.6}\\
 $\Rm$ & $U a/ \eta$ &&\multicolumn{2}{c}{0.55}& \multicolumn{2}{c}{44}\\
 $\Re$ & $U a/ \nu$ &&\multicolumn{2}{c}{$7\;10^4$}& \multicolumn{2}{c}{$6\;10^6$}\\
 & & & $\underline{\: B=B_i \:}$ & $\underline{\: B=B_o \:}$&$\underline{\: B=B_i \:}$ & $\underline{\: B=B_o \:}$\\
 $\N$ & $\sigma a B^2/\rho U$ & & 262 & 0.06& 3.1& 0.01\\
 \hline
\vspace{0.2cm}
\end{tabular}

\end{center}
\caption{Typical values of the relevant parameters and dimensionless numbers for different imposed rotation frequencies $f = \Omega/2\pi$ of the outer sphere and differential rotation frequencies $\Delta f= \Delta\Omega/2\pi$ of the inner sphere. For the numbers that depend on the magnetic field strength, two values are given, the first one with $B=B_i=0.175$ T at the equator of the inner sphere, the second one with $B=B_o=0.008$ T at the equator of the outer sphere.}
    \label{tab:dimen}
\end{table}
\section{Results}
\label{Results}

\subsection{A typical run}

\begin{figure}
  \centerline{ \includegraphics[width=13cm]{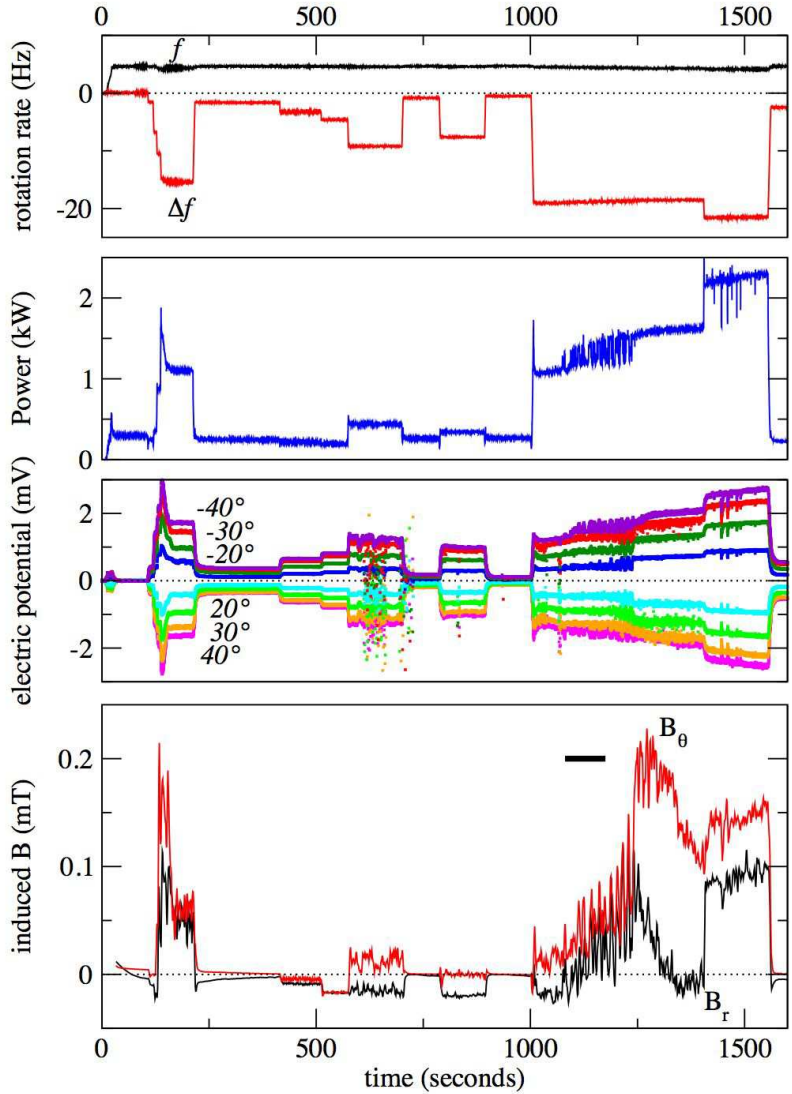}}
     \caption{An example of the data acquisition during an experimental run, as a function of time (in seconds). The top graph
shows $f$, the rotation rate of the outer sphere, and $\Delta f$, the rotation rate of the inner sphere with
respect to the outer sphere. The next graph below is the total power. The differences in electric potential
recorded by the eight longitudinal electrode pairs (from $-40\deg$ to $40\deg$ latitude) is shown next. The last graph below gives the $B_r$ and $B_{\theta}$ components of the induced magnetic field (in mT) recorded on the $50\deg$-latitude magnetometer, corrected for temperature. The horizontal
bar in that graph gives the time window of the angular velocity record shown in figure \ref{fig:dop}.}
        \label{fig:run_type}
\end{figure}

Figure \ref{fig:run_type} illustrates the data typically acquired during a 1600s-long run. The top graph shows
the variation of $f$, the rotation frequency of the outer sphere, and $\Delta f$, that of the inner sphere with respect to the outer sphere. The outer sphere is first set to a rotation rate of about 5 Hz, then several plateaux of different $\Delta f$ are produced, down to $\Delta f$ = -21.5 Hz. Here, they all correspond to counter--rotation. Some small fluctuations of the rotation rates are observed ($\pm 3\%$ max for $\Delta f$ and $\pm 7\%$ max for $f$).
They are due to variations in the power demanded to the motors, as shown in the next graph below, which displays the
total power of the motors, reaching 2.3 kW. Note the large power fluctuations ($\pm 22\%$) in the middle of the plateau with $\Delta f$ = -19 Hz.
The next graph gives the differences in electric potential $\Delta V$ recorded by the eight electrode pairs along a meridian.
The highest signal reaches 3 mV. As expected from equation \ref{eq:uphi}, the potential difference is largest at the highest
latitude because $B_r$ is largest there. Since $\Delta f$ is negative, the electric potential differences are positive in the
southern hemisphere and negative in the northern hemisphere (with our connection conventions). There is an almost perfect symmetry between the two hemispheres.
Several ultrasonic Doppler velocity profiles are recorded during this sequence. The
last graph below shows the induced magnetic field measured at $50\deg$ latitude. The records have been time-averaged along
a 3 seconds--window to remove the fluctuations linked with the rotation of the outer sphere,
and the imposed dipole contribution has been removed. The plateaux are well identified, but large
fluctuations are observed when $\Delta f$ is large. The maximum amplitude of the induced magnetic field is 0.23 mT ($2.4\%$ of the
imposed field at this position).

\begin{figure}
  \centerline{ \includegraphics[width=13cm]{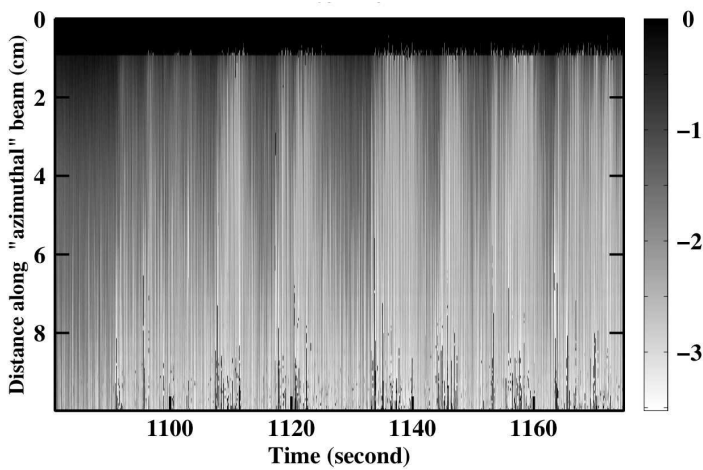}}
     \caption{An example of an ultrasonic Doppler angular velocity acquisition.
The horizontal axis is time (in seconds) and the vertical axis is distance along the shooting line (in cm),
here down to $d \simeq 10$ cm. The corresponding time--window is drawn in figure \ref{fig:run_type}.
The shades give the velocity value in m/s (scale on the right-hand side). Strong velocity fluctuations are observed.}
        \label{fig:dop}
\end{figure}

A typical ultrasonic Doppler velocity record is shown in figure \ref{fig:dop}. It corresponds to the time window marked with
a horizontal bar in the last graph of figure \ref{fig:run_type}, when time--oscillations are observed on all data. The spatiotemporal
plot clearly shows these oscillations. The velocities are largest at the bottom of the figure, where the ultrasonic beam gets closest to the inner sphere for
this record. Equation \ref{eq:sd} gives the value of the cylindrical radius $s = 12.1$ cm. 
There, the velocity reaches -3.5 m/s. This large value is nevertheless smaller than the imposed tangential velocity on the inner sphere (-8.7 m/s). For this file, the time--interval between profiles is 15 ms, and there are
128 shots per profile. Velocities cannot be measured in the first 1 cm of the records, due to strong
reverberations in the assembly wall.

Note that all data display large fluctuations in the time--window between 1100 and 1250 seconds. We will see that, in this parameter range, the flow exhibits a fundamental transition between two different configurations (see section \ref{Pic_Nadege}). Here, this transition is preceded by large oscillations, while in some other runs, it occurs in one single step.

\subsection{Evolution with the imposed differential rotation}

\begin{figure}
  \centerline{ \includegraphics[width=11cm]{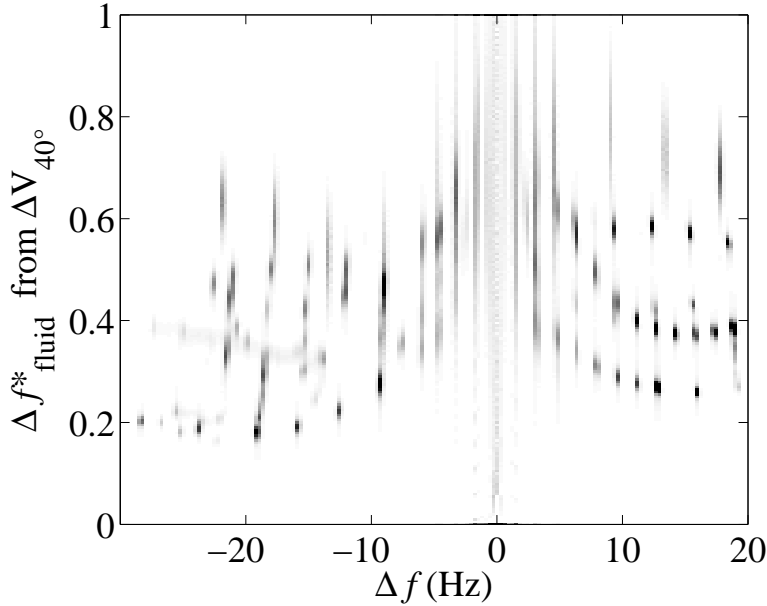}}
     \caption{A compilation of the variation of the normalized angular velocity of the liquid as a function
of the imposed differential rotation rate $\Delta f$. The velocity of the outer sphere is $f \simeq 4.5$ Hz for
all data shown. The shades give the proportion of data points that yield a given couple of values. Note
that very different angular velocities are measured for a given $\Delta f$.
}
        \label{fig:ddp40_vs_df}
\end{figure}
\begin{figure}
  \centerline{ \includegraphics[width=11cm]{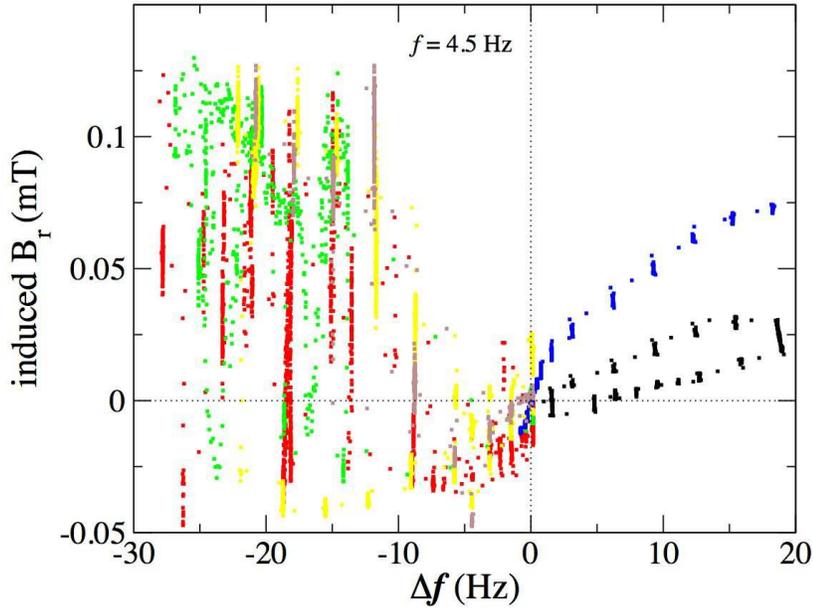}}
     \caption{A compilation of the variation of the radial component of the induced magnetic field as a function
of the imposed differential rotation rate $\Delta f$. The velocity of the outer sphere is $f \simeq 4.5$ Hz for
all data shown. The colors indicate six different experimental runs. Note the large scatter of the data points,
especially for negative $\Delta f$.}
        \label{fig:Br_vs_df}
\end{figure}

Figure \ref{fig:ddp40_vs_df} shows the dimensionless angular frequencies $\Delta f^*_\text{fluid} = {U_{\varphi}}/{2 \pi s \Delta f}$ deduced from
the electric potential differences measured at $40\deg$ latitude (according to equation \ref{eq:uphi}), as a function of the imposed differential rotation $\Delta f$, for
a given rotation rate $f \simeq 4.5$ Hz of the outer sphere. The figure is similar to figure 4 of \cite{Nataf06}, but
data from several additional runs are included. The data points appear very scattered: for a given
$\Delta f$, the dimensionless angular velocity can vary by up to a factor of 3. We think that this is due to variations in the electric coupling between liquid sodium and the inner copper sphere. The coupling between the rotating
inner sphere and the liquid sodium depends on the electric currents that are induced in the sodium and loop
in the copper inner sphere. 
A similar scatter is observed when plotting the induced magnetic field as a function of $\Delta f$ (figure \ref{fig:Br_vs_df}). 

This scatter makes it difficult to draw robust conclusions about features such as super--rotation, but we will show in the next section that there is a linear relationship between the electric
potential difference at $40\deg$ ($\Delta V_{40\deg}$) and the azimuthal velocities in the sodium actually measured with ultrasonic Doppler
velocimetry. We will therefore use $\Delta V_{40\deg}$ as a proxy of the actual flow velocity, and we will see
that several features of the flow become coherent when this proxy is used in place of $\Delta f$.

\subsection{Angular velocity profiles}

\begin{figure}
  \centerline{ \includegraphics[width=10cm]{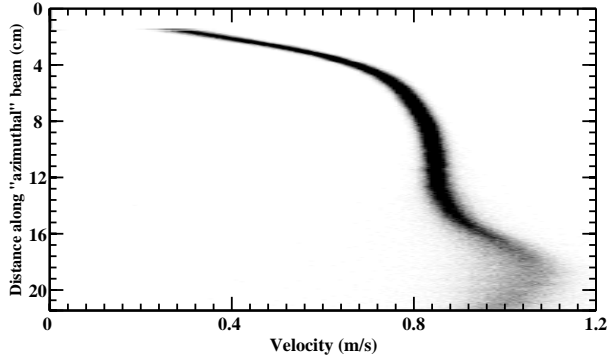}}

     \caption{Angular velocity profile for $f = 4.7$ Hz and $\Delta f = 4.8$ Hz. The time--interval between profiles
is 30 ms, and there are 128 shots per profile. Probability density function or histogram of the velocity as a function of distance along shot.}
        \label{fig:dop_profile}
\end{figure}

\begin{figure}
  \centerline{ \includegraphics[width=10cm]{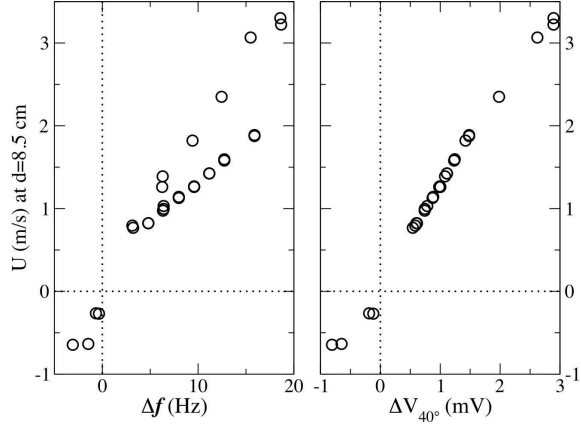}}
     \caption{Velocity (in m/s) at a distance of 8.5 cm along the shot line ($s \simeq 13.6$cm), as a function of $\Delta f$ (left) and $\Delta V_{40\deg}$ (right). The rotation rate of the outer sphere is $f \simeq 4.5$ Hz.
The linear relation we get on the right shows that $\Delta V_{40\deg}$ is a good proxy of the angular velocity of the
fluid at depth.}
        \label{fig:uphi_vs_ddp40}
\end{figure}

Using Doppler ultrasonic velocimetry, we are able to measure angular velocity of the liquid sodium flow. Figure \ref{fig:dop_profile} shows a profile of velocities measured along the shooting line for $f = 4.7$ Hz and $\Delta f = 4.8$ Hz.
Equation \ref{eq:umes} shows that the measured velocity component is proportional to the angular velocity of the fluid.
Actually, the plot is built from a 40 s--long record and the shades map the probability density
function ($pdf$) of velocity versus distance.
In this regime,
the flow is almost steady and the $pdf$ is sharp. Velocity
rises steadily from a value close to 0.3 m/s at the outer boundary (top of the graph) to reach a plateau of about 0.85 m/s. The velocity
further increases to a bump at 1.05 m/s where the ultrasonic beam gets closest to the inner sphere ($d \simeq 18.8$ cm, $s \simeq 7.9$ cm). 
We interpret this bump as the signature of the magnetic wind, which is confined to the neighborhood of the inner sphere,
where the magnetic field is largest. The plateau characterizes the region where the fluid is entrained at an almost
uniform angular velocity. This profile is typical of
those measured when the outer sphere is rotating, and differs from those obtained when the outer sphere is at rest.
For all available profiles, we pick the velocity measured at a distance of 8.5 cm (on the plateau, and about
at mid--depth of the fluid layer, in the equatorial plane, according to figure \ref{fig:numerics}) and plot it as a
function of the corresponding $\Delta f$ and $\Delta V_{40\deg}$ (Figure \ref{fig:uphi_vs_ddp40}). While
the former plot displays a large scatter, all points align well when velocity is plotted against $\Delta V_{40\deg}$, illustrating
the idea that the latter represents a good proxy of the effective velocity. From the regression line, we can
translate $\Delta V_{40\deg}$ into fluid velocity or fluid flow frequency at depth in the fluid. We obtain: ${\Delta f_\text{fluid}}/{\Delta V_{40\deg}} \simeq 2600$ Hz/V. Note that the fluid velocity
deduced from $\Delta V_{40\deg}$ using equation \ref{eq:uphi} agrees with the azimuthal velocity measured by ultrasonic Doppler velocimetry at the cylindrical radius corresponding to this latitude
of $40\deg$.

\begin{figure}
  \centerline{ \includegraphics[width=12cm]{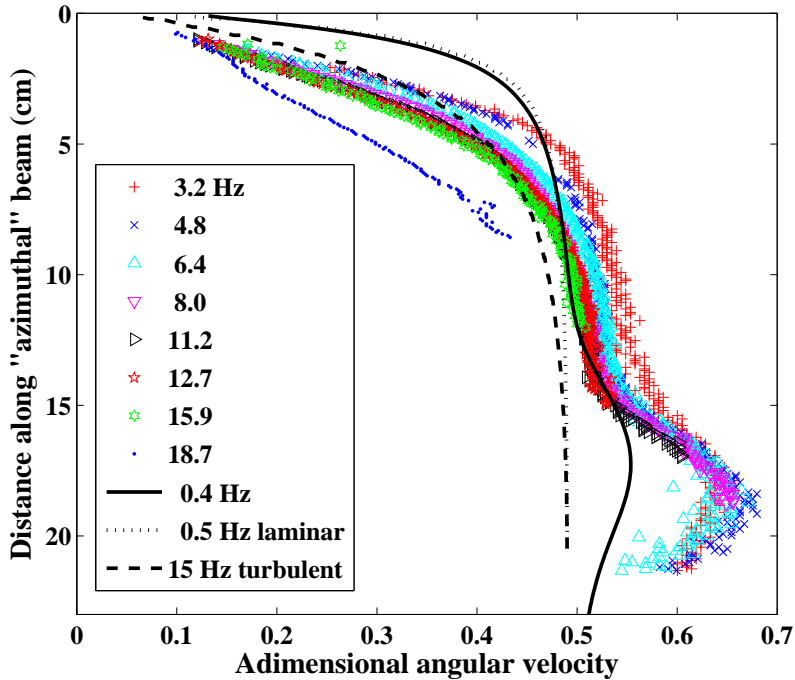}}
     \caption{A compilation of the angular velocity profiles, normalized by $\Delta V_{40\deg}$, for
$f \simeq 4.5$ Hz and for imposed $\Delta f$ ranging from 3.2 to 18.7 Hz. The horizontal scale has been chosen
to give the actual adimensional angular velocity for the reference profile with $\Delta f = 3.2$ Hz. The continuous lines
are predictions of different sort: the solid line is the profile obtained from the numerical model of figure \ref{fig:numerics},
 ($\Delta f = 0.4$ Hz) with the electrical conductivity of the inner sphere reduced by a factor 100 (see section \ref{variable-coupling}).
The almost superposed dotted line is derived from the analysis of \cite{Kleeorin97}, and the dashed line is the prediction
of our extended torque--balance model for $\Delta f = 15$ Hz (see section \ref{torque-model}). The two latter models only predict
the shape of the boundary--layer profile, not the height of the plateau of near solid--body rotation.
}
        \label{fig:uphi_profile}
\end{figure}

In figure \ref{fig:uphi_profile}, we have plotted all the profiles for $f \simeq 4.5$ Hz and positive $\Delta f$, divided by the corresponding 
$\Delta V_{40\deg}$. The horizontal scale has been chosen to provide the actual adimensional angular velocity
for the reference case $\Delta f = 3.2$ Hz, according to equation \ref{eq:umes}.
However, one should not consider this scale as universal, since the actual coupling is very variable (see figure \ref{fig:ddp40_vs_df}). We have also drawn a synthetic profile derived from a numerical simulation similar to the one shown in figure \ref{fig:numerics} with $\Delta f = 0.4$ Hz, but with the electric conductivity of the inner
sphere reduced from its nominal value of 4.2 times that of the liquid sodium down to 0.04.
In a first approximation, all experimental profiles collapse on a single curve,
showing that the normalization with $\Delta V_{40\deg}$ is adequate. A closer look reveals that the rise of
the velocity from the outer boundary is sharper when the forcing is weaker. In other words, the plateau of
high angular velocity extends further towards the outer sphere for weak forcing. 
We will present in section \ref{torque-model} a simple model, in the spirit of \cite{Kleeorin97}, that reproduces these characteristics (dotted and dashed lines).
It seems that the relative size
of the bump due to the magnetic wind above the plateau does not depend much upon the forcing.
The comparison with the numerical profile suggests that the zone of strong magnetic wind is closer to the inner sphere
in the experiments than in the simulation.
We also note that the profile
that corresponds to the highest differential rotation rate $\Delta f = 18.7$ Hz plots below the others and has a
distinct kink near the outer boundary, which makes it look more alike the profiles obtained with $f = 0$.

\subsection{Electric potentials}

\begin{figure}
  \centerline{ \includegraphics[width=11cm]{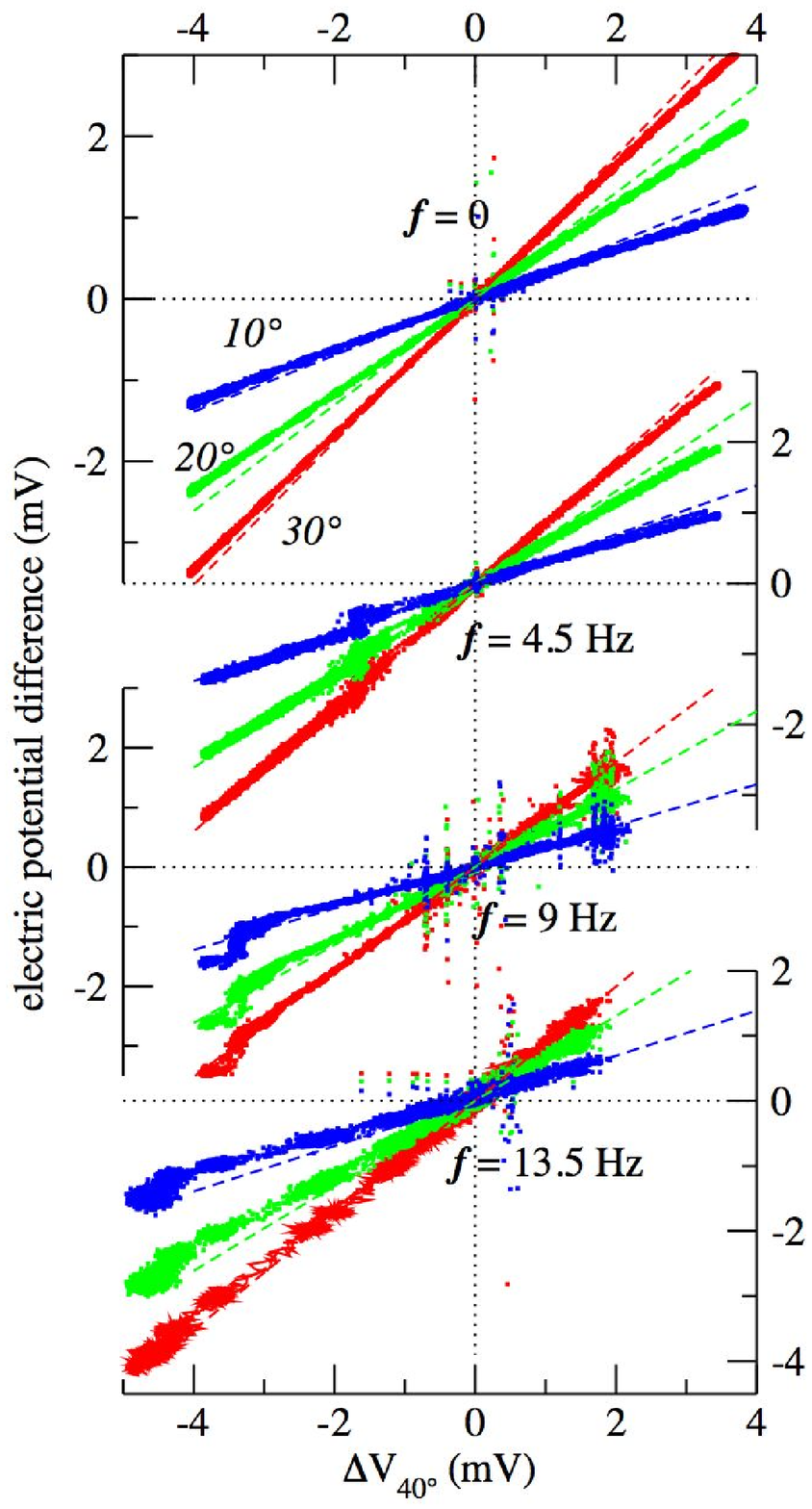}}
     \caption{A composite of electric potential differences for four different rotation rates $f$ of the outer sphere.
The electric potential difference measured at latitudes of $10\deg$, $20\deg$, and $30\deg$ are plotted against that measured at $40\deg$.
The relation is almost linear, except for a kink observed around a negative value of $\Delta V_{40\deg}$, which moves towards
more negative values when $f$ increases. The dashed straight lines indicate the relationship one would get if the fluid was in solid body rotation. Scattered points in the bottom panels are due to electromagnetic noise.}
        \label{fig:ddp_vs_ddp40}
\end{figure}
We showed in \cite{Nataf06} that, when the outer sphere is at rest, the relative variation of electric potential differences (normalized to its value at $40\deg$) 
with latitude is independent of the differential rotation rate $\Delta f$. The better data now available confirm this
observation and further demonstrate
that the latitudinal variation of potentials is close to, but significantly different from, the one expected for solid body rotation.
Figure \ref{fig:ddp_vs_ddp40} gives the variation of electric potential differences $\Delta V$ at latitudes of $10\deg$, $20\deg$ and
$30\deg$ as a function of $\Delta V_{40\deg}$, for three different rotation rates of the outer sphere ($f \simeq 4.5, 9$ and 13.5 Hz), and several different $\Delta f$ (both positive and negative). The case with no rotation ($f = 0$) is included for reference.
In all four cases, the data points align well, and fall not far from (but distinctly off) the straight lines calculated for solid body rotation ({\it i.e.}, assuming $U_{\varphi}$ proportional to $s$ in equation \ref{eq:uphi}). In fact, we deduce from the measured trend for $f =4.5$ Hz that the angular velocity at $30\deg$
is about 0.9 times that at $40\deg$, in good agreement with the angular velocity profiles of
figure \ref{fig:uphi_profile}. At lower latitudes, the apparent velocities deduced from the electric
potentials are larger than the actual velocities, as the assumptions used in equation \ref{eq:uphi}
break down \citep{Nataf06}.
In contrast with this simple linear trend, there is a distinct interval of $\Delta V_{40\deg}$ that displays a peculiar behavior, when the outer sphere is rotating. It occurs around $\Delta V_{40\deg} = -1.7$ mV for $f=4.5$ Hz, -3.3 mV for $f=9$ Hz, and -4.4 mV for $f=13.5$ Hz. 
The kink means that, there, the variations in electric potentials are larger at low latitudes than at $40\deg$.
We will see below that the induced magnetic field
also displays a very peculiar variation around the same values of $\Delta V_{40\deg}$.

\subsection{Induced magnetic field}

\begin{figure}
  \centerline{ \includegraphics[width=8cm]{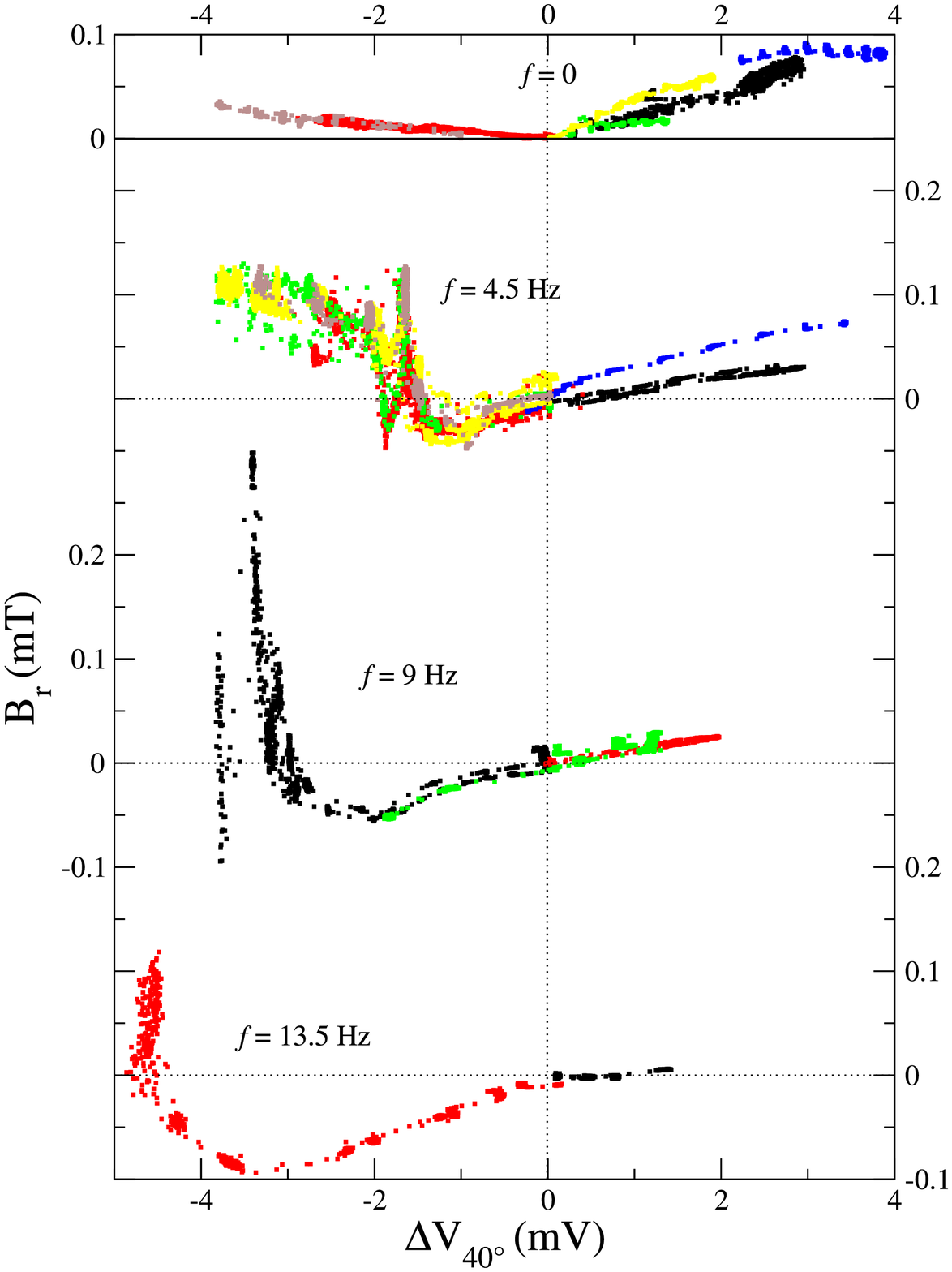} \hspace{0.5cm} \includegraphics[width=8cm]{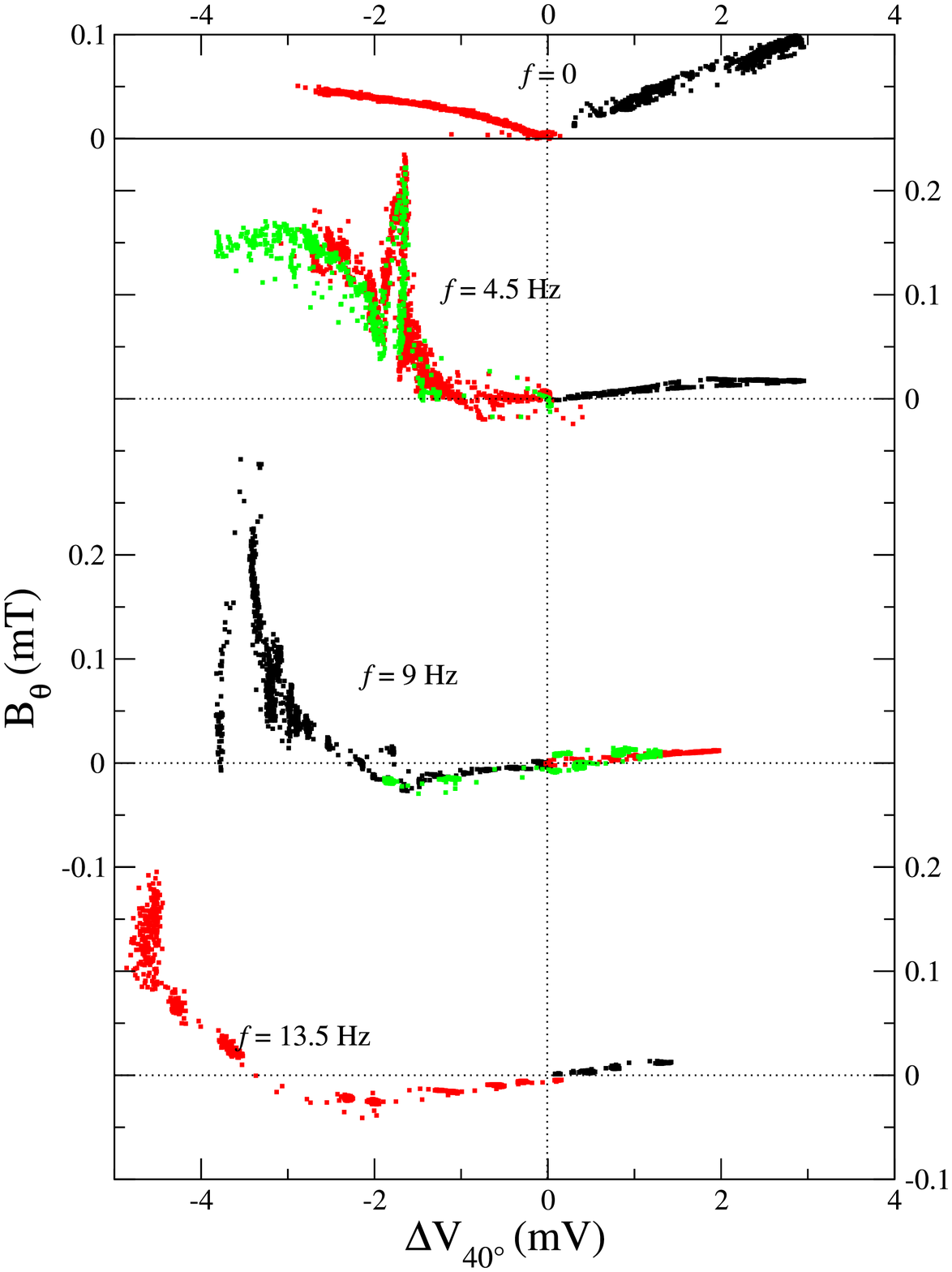}}
\vspace{1.5cm}
     \caption{A composite of the induced magnetic field as a function of $\Delta V_{40\deg}$, for four different rotation frequencies $f$ of
	 the outer sphere. Radial component on the left, and orthoradial component on the right. Note the symmetry change around the origin ($\Delta f \simeq 0$) between the case without rotation ($f=0$)
and the other cases. A striking peak is observed for a particular value of $\Delta f$ in the counterrotating
regime ($\Delta V_{40\deg} < 0$). The colors correspond to different runs.}
        \label{fig:B_vs_ddp40}
\end{figure}

The induced magnetic field is measured outside the sphere, in the laboratory frame. The mean axisymmetric part of the induced field
results from the interaction of the imposed axial dipolar magnetic field with the mean axisymmetric part of the meridional velocity field.
The magnetometer signals are first averaged over 3 seconds, in order to remove the oscillatory part (which is discussed in another article \citep{Schmitt08}) and
other magnetic perturbations. We then correct for temperature drift, by removing an empirical linear correction. In figure \ref{fig:B_vs_ddp40}, both the radial $B_r$ and orthoradial $B_{\theta}$ components
are plotted as a function of $\Delta V_{40\deg}$ for several different imposed rotation frequencies $f$ of the outer sphere.
We have included the $f=0$ case for reference. 

The first thing to note is the change of symmetry about $\Delta V_{40\deg} = 0$ for small forcing : while the
induced magnetic field is always positive when the outer sphere is at rest, we find that the induced field has the sign of $\Delta V_{40\deg}$ (or $\Delta f$) when the outer sphere is rotating. This can be
directly related to the sense of the meridional circulation. When the outer sphere is at rest, the flow is always centrifugal in the equatorial plane, irrespective of the
sign of $\Delta f$. 
When the outer sphere is rotating, the sense of the mass flux through the outer viscous boundary layer is governed by the sign
of the differential rotation between the fluid and the outer sphere (as for Ekman pumping). When $\Delta f/f$ is negative
and small enough, the mass flux out of the boundary layer is thus centripetal at the equator.

In that same regime, we also note that the induced magnetic field is weaker when the outer sphere is rotating : the meridional circulation violates the
Proudman--Taylor constraint and therefore decreases as the Ekman number decreases (see figure 6 of \cite{Hollerbach07}).

However, the most interesting feature of the curves shown in figure \ref{fig:B_vs_ddp40} is the distinctive narrow peak observed at negative $\Delta f$ (here negative  $\Delta V_{40\deg}$).
The peak occurs for a value of $\Delta V_{40\deg}$ that moves toward more negative values as $f$ increases. In fact, the occurrence of the
peak corresponds to the transition between two different regimes, which was invoked when discussing figure \ref{fig:ddp_vs_ddp40}. We stress that
it is because we plot the induced magnetic field as a function of $\Delta V_{40\deg}$ rather than $\Delta f$ that we obtain a well defined and reproducible
peak (compare with figure \ref{fig:Br_vs_df}). The peak happens as the flow evolves from centripetal to centrifugal in the equatorial plane, as the effect of differential rotation takes over the
effect of global rotation.

\subsection{Radial velocity profiles}

\begin{figure}
  \centerline{ \includegraphics[width=8cm]{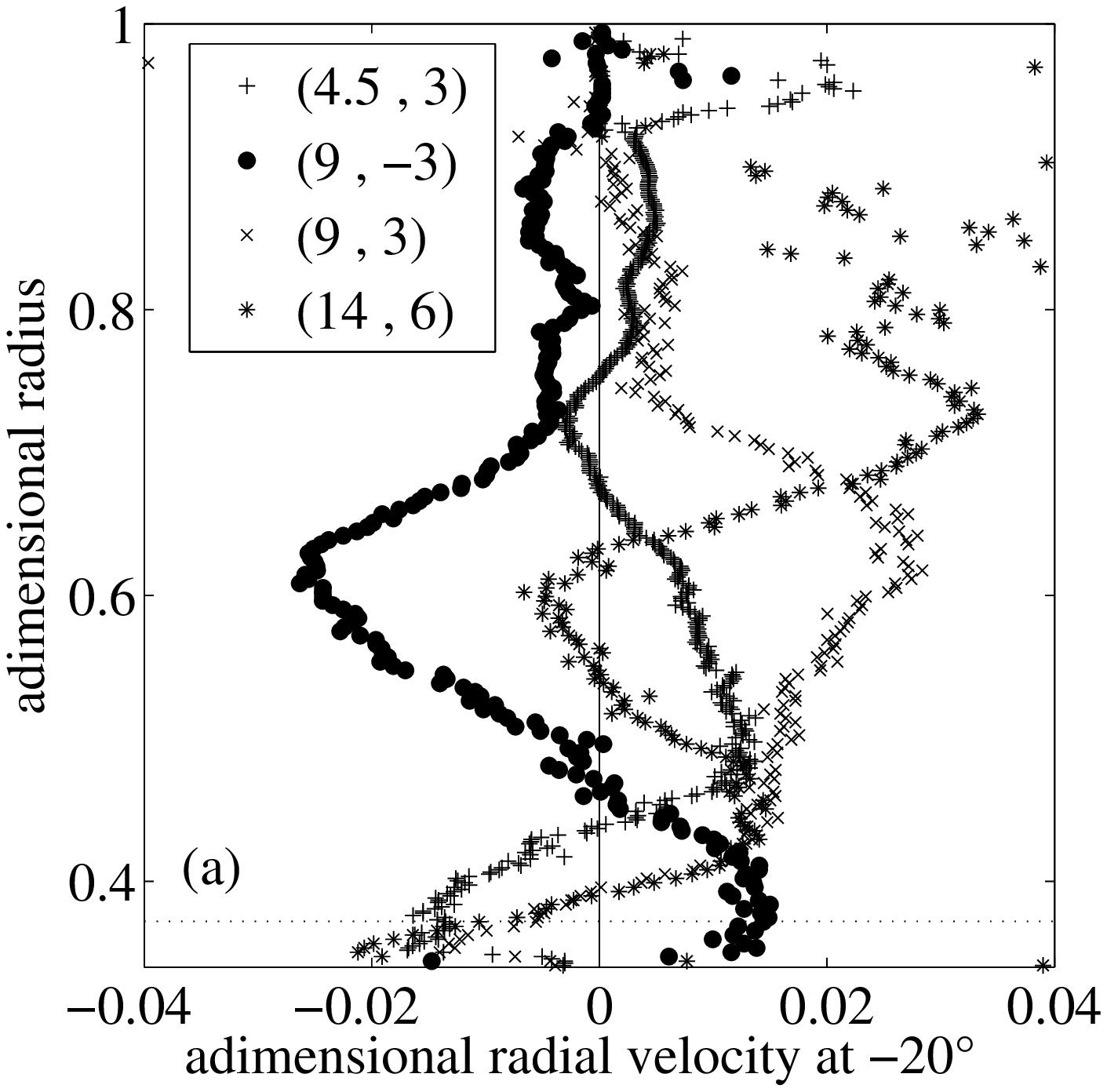} \hspace{-0.5cm} \includegraphics[width=8cm]{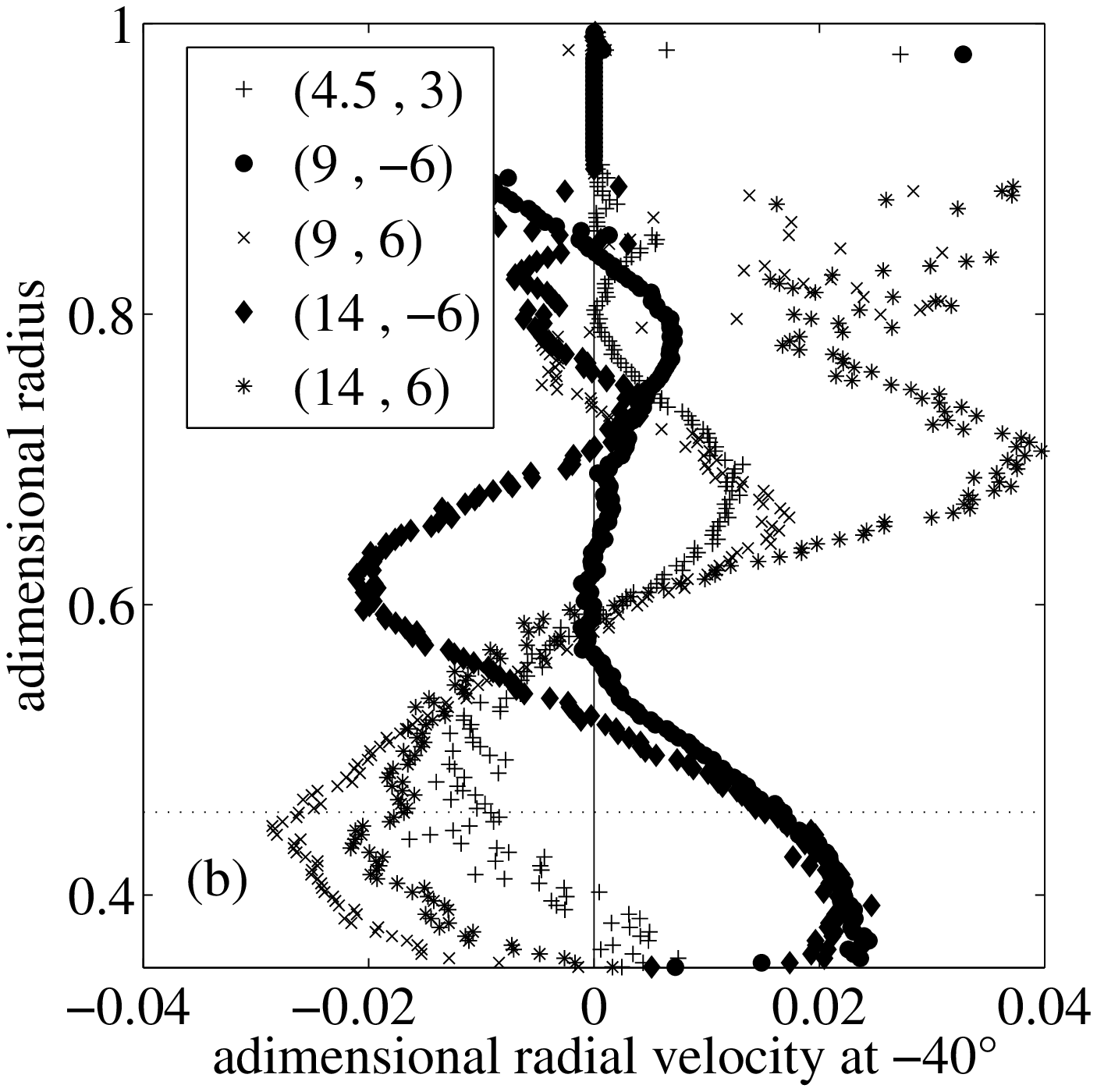}}
     \caption{Radial velocity, obtained by ultrasonic Doppler velocimetry, as a function of adimensional radius (vertical axis) at two
different latitudes: $-20\deg$ (a) and $-40\deg$ (b). The dotted line indicates the radius of the tangent cylinder
on the inner sphere, whose adimensional radius is 0.35 (bottom of the plots). All velocities are normalized with
the differential tangential velocity at the equator of the inner sphere. The profiles are labelled by their ($f$, $\Delta f$) pair.
Note
that the sign of the radial velocity changes as the sign of $\Delta f$, as expected for Ekman--pumping dominated flow.}
        \label{fig:Ur}
\end{figure}
While the induced magnetic field measured outside the sphere is a good indicator of the global symmetries of the meridional flow,
it cannot be used
to investigate the internal organization of that flow. Here, we add the information brought by radial profiles
of the radial velocity obtained by ultrasonic Doppler velocimetry. A compilation of the profiles obtained for three different rotation frequencies $f$ of the outer sphere is
given in figure \ref{fig:Ur} for two different latitudes: $-20\deg$ (figure \ref{fig:Ur}a) and $-40\deg$ (figure \ref{fig:Ur}b). The profiles clearly confirm that the sign of the radial velocity changes with the sign of $\Delta f$, as
inferred from the induced magnetic field measurements.
At $-40\deg$, the profiles are characterized by a broad peak with radial velocity rising from zero at the inner sphere to a
maximum, which can reach about $3\%$ of the tangential velocity of the inner sphere. There, the flow is centripetal when
the Rossby number is positive. The radial velocity falls back to zero at nearly the same radius for all $f$, but the profiles
for the higher $f$ (lower Ekman number) are characterized by a strong second peak.
At $-20\deg$, the amplitudes are similar but seem to increase with $f$. The change of sign as $\Delta f$ changes sign is
well visible again.

\section{Discussion and conclusions}
\label{Discussion}

\subsection{Moderate--Rossby number regime}

The magnetostrophic regime at moderate Rossby numbers is characterized by a low level of turbulence. Most of the energy is
in the axisymmetric mean flow described in this article. The meridional circulation has the symmetry of Ekman pumping:
counterclockwise in a meridional plane ({\it i.e.}, centrifugal at the equator) for $\Ro > 0$ and clockwise for $\Ro < 0$. Radial velocities are at least one order of magnitude smaller
than azimuthal velocities. Waves are observed in this regime, and are described in
another article \citep{Schmitt08}.

Profiles of the angular velocity of the flow have been obtained, using ultrasonic Doppler velocimetry. They
reveal a region where magnetic wind is present, near the inner sphere, where the magnetic field is largest. Farther
away, the flow is geostrophic, under the effect of the Coriolis force. A large part of the geostrophic region is
entrained by the magnetic wind at a nearly constant angular velocity.

The measurements show that the shape of the
angular velocity profile is nearly independent of $\Ro$ as long as it remains moderate. This suggests that, in contrast
to the case with a fixed outer sphere \citep{Nataf06, Hollerbach07}, the linear solution ($\Ro << 1$) provides a
good description of the flow. We were able to compute solutions in the linear regime with the geometry and parameters
of the $DTS$ experiment for $f = 5$ Hz ($\E = 4.7 \, 10^{-7}$), as shown in figure \ref{fig:numerics}. 
In figure 
\ref{fig:numerics}a, one clearly sees a non--geostrophic inner region where the Ferraro law of isorotation applies, a central
part where the flow is geostrophic and in almost solid--body rotation, and an outer domain, which is also geostrophic but
in which the angular velocity decreases rapidly towards the outer sphere. 
In order to
compare with the experimental results, synthetic velocity profiles have been computed.
The comparison of the angular velocity profiles is shown in figure \ref{fig:uphi_profile}.
All profiles show the plateau of nearly constant angular velocity, the central peak due to the magnetic wind, and the
gentle drop to lower velocity towards the outer sphere. We note however that this velocity drop is much steeper in the numerical
simulation (which has $\Ro=0.08$). Several of these features had been predicted by \cite{Kleeorin97} in an
asymptotic study of spherical Couette flow in a dipolar magnetic field. We recall their main results below
and propose an extension of their approach, which better explains our experimental results.

\subsection{Modified Taylor constraint and extended quasi--geostrophic model for the moderate--Rossby number flow}
\label{torque-model}
The quasi--geostrophic regime with an imposed magnetic field has been studied extensively by \cite{Kleeorin97}. Following the pionneer computations of \cite{Hollerbach94} of rotating spherical Couette in a dipolar magnetic field, they worked out the asymptotic solutions for small Elsasser, Ekman, Reynolds and Rossby numbers. They predicted the existence of a geostrophic region where the magnetic torque on
coaxial cylindrical tubes balances weak friction in the upper and lower Ekman layers, thus providing an extension of
the zero--Ekman number analysis of \cite{Taylor63}, which results in the so--called Taylor--state. We consider
the torques at work on a cylindrical tube at cylindrical radius from $s$ to $s+ds$. Neglecting the volumetric viscous friction
(small Ekman number), the torque balance can be written as:

\begin{equation}
d\gamma_T(s) + d\gamma_B(s) + d\Gamma_M(s) = 0,
  \label{eq:torque_balance}
\end{equation}

where $d\gamma_T$ and $d\gamma_B$ are the (resistive) torques on the top and bottom spherical caps at the outer sphere, and $d\Gamma_M$
is the volumetric (driving) torque exerted by the Lorentz force (magnetic torque). In our case, the surface torques are due to Ekman
friction on the outer sphere and can be written:

\begin{equation}
d\gamma_T(s) = d\gamma_B(s) = 2 \pi s^2 \tau_V \frac{ds}{\cos \theta},
  \label{eq:Ekman_torque}
\end{equation}

where $\theta$ is the colatitude and $\tau_V$ is the azimuthal component of the friction stress on the outer shell.

At high latitudes, friction in the Ekman layers is weak and, as in the Taylor state, the balancing magnetic torque vanishes in the steady--state
and the fluid is in solid body rotation. The magnetic torque can be expressed as:

\begin{equation}
d\Gamma_M(s) = 2 \pi s^2 ds \int_{z_B}^{z_T}{(\vec{j} \times \vec{B})_{\varphi} \, dz} = \frac{2 \pi}{\mu_0} ds \frac{d}{ds} \left[s^2 \int_{z_B}^{z_T}{B_s b_{\varphi} \, dz} \right],
  \label{eq:magnetic_torque}
\end{equation}

where $B_s$ is the component of the imposed magnetic field along the cylindrical radius $s$, while $b_\varphi$ is the (unknown) azimuthal component of the induced magnetic field. This component results from the shearing of the dipolar field lines by deviations of the fluid flow from solid body rotation.
A weak magnetic torque thus implies a small deviation from solid body rotation, especially where the imposed field $B_s$ is large, {\it i.e.} near the inner sphere (but not too close, because the flow is no longer geostrophic there). As noted by \cite{Kleeorin97},
this explains why the angular velocity of the fluid $\Delta \omega(s)$ is almost constant in the central part of the spherical shell.
The progressive decrease of the angular velocity as $s$ increases further was predicted by \cite{Kleeorin97}. In this layer,
which they call ``magnetic--Proudman'', shear in the upper and lower Ekman layers increases, so as to produce a large enough $b_\varphi$ to compensate
for the decrease of $B_s$ in the expression of the magnetic torque. In order to be more quantitative, one needs to express
$b_\varphi (s,z)$ as a function of $\Delta \omega(s)$. \cite{Kleeorin97} show that, when the imposed magnetic field is a dipole,
the magnetic torque can be written:

\begin{equation}
d\Gamma_M(s) = \frac{12 \pi}{\mu_0 \eta} B_0^2 a^5 \Delta\Omega  \frac{ds}{a} \left[2\left(1-\frac{s}{a}\right) \right]^{3/2} \frac{\Delta\omega^{\dag} - \Delta\omega(s)}{\Delta\Omega},
  \label{eq:magnetic_torque_2}
\end{equation}

where $\Delta \omega^{\dag}$ is the differential rotation of the fluid in the region of nearly solid--body rotation. The friction
stress for the laminar Ekman layer is easily calculated and yields:


\begin{equation}
d\gamma_T(s) = - 2 \pi \rho a^4 \Delta\Omega \sqrt{\nu \Omega} \; \frac{ds}{a} \frac{s^3}{a^3} \left[ 1-\frac{s^2}{a^2}\right]^{-1/4}  \frac{\Delta \omega(s)}{\Delta \Omega}.
  \label{eq:Ekman_torque_laminar}
\end{equation}

Injecting expressions \ref{eq:magnetic_torque_2} and \ref{eq:Ekman_torque_laminar} in equation \ref{eq:torque_balance}
in the boundary layer approximation ($s \simeq a$), \cite{Kleeorin97} obtain the shape of the profile of the angular velocity versus $s$:

\begin{equation}
\Delta \omega(s) = \frac{\Delta \omega^{\dag}}{1 + \frac{\rho \sqrt{\nu \Omega}}{3 a \sigma B_0^2} \left[ 2 \left( 1- \frac{s}{a} \right) \right]^{-7/4}}
  \label{eq:Kleeorin_profile}
\end{equation}

The synthetic angular velocity profile derived from this expression is the dotted line of figure \ref{fig:uphi_profile}. The value
of $\Delta \omega^{\dag}$ has been chosen to coincide with the full numerical profile drawn (solid line).
Indeed, \cite{Kleeorin97} studied the case when the flow remains geostrophic all the way to the inner sphere, where they
matched the solid--body solution to the Stewartson layer modified by the magnetic field. They could thus predict the value of
$\Delta \omega^{\dag}$. In our case, the inner region is not geostrophic, and we only use their analysis to probe the
shape of the angular velocity profiles in the geostrophic region.
It is clear
that, in that region, the asymptotic prediction of \cite{Kleeorin97} is a very good approximation of the actual numerical profile.
However, we note that both profiles fail to explain the much more gentle rise of angular velocity observed in the experiments.
The explanation is that the Ekman boundary layer becomes turbulent at the high $\Delta f$ reached in our experiments.
We can retain the approach of \cite{Kleeorin97} and simply allow for turbulent friction stress $\tau_V$
in equation \ref{eq:Ekman_torque}. Because of turbulence, Ekman friction is enhanced and a stronger magnetic torque is needed to balance it. This results in a
widening of the ``Proudman--magnetic'' layer, as shown by the dashed line in figure \ref{fig:uphi_profile}, computed
for $\Delta f = 15$ Hz. The widening
increases as $\Delta f$ increases, in excellent agreement with our measurements. The full ingredients and implications of
this extended model will be given elsewhere.


\begin{figure}
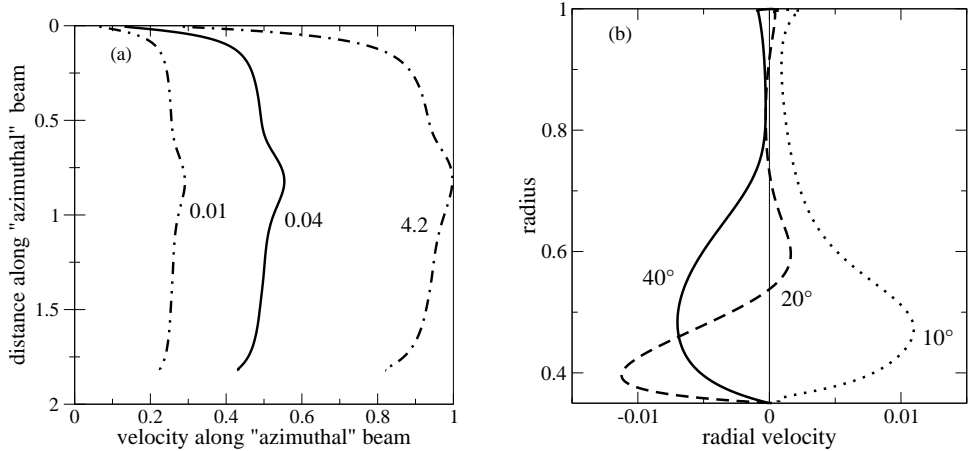

  \centerline{ \includegraphics[width=6cm]{fig13a.eps} \hspace{0.5cm} \includegraphics[width=6cm]{fig13b.eps}}
     \caption{Synthetic Doppler velocity profiles computed from numerical simulations.
(a) angular velocity synthetics for a shot from
the azimuthal assembly at $10\deg$ (see projected raypath in figure \ref{fig:numerics}a). The synthetics take into
account the finite width of the ultrasonic beam and the projection of the meridional velocity. The dash--dot line is for the numerical simulation of figure \ref{fig:numerics}, in which
the electric conductivity of the inner sphere is 4.2 times that of the liquid sodium, as expected for copper. The other computations have a reduced inner sphere conductivity ratio of 0.01 (dot--dot--dash), and 0.04 (solid line). The corresponding angular velocity profile for the latter is also drawn in figure \ref{fig:uphi_profile}. (b) radial
velocity synthetics for radial shots at latitudes $\pm 40\deg$, $\pm 20\deg$ and $\pm 10\deg$ for the latter model.
Velocities are normalised with the differential tangential velocity at the equator of the inner sphere, and distances are normalised by the radius of the outer sphere.
}
        \label{fig:synthetics}
\end{figure}

\subsection{Variable electric coupling and $\Ro_{eff}$}
\label{variable-coupling}

In \cite{Nataf06}, there were hints that different flows could be obtained for the same imposed rotation rates $f$ and
$\Delta f$. Our new data confirm this trend. We speculate that this behavior is due to variations in the electric coupling between liquid sodium
and the copper inner sphere, under the effect of electrowetting feedback. 
It is thus difficult to relate the observed
signals to the imposed $\Delta f$, as we would like to do in order to study the possibility for the magnetic wind
to induce super--rotation of the fluid as in \cite{Dormy98}. However, we find that most results can be cast in a coherent frame if we use
$\Delta V_{40\deg}$, the electric potential difference at latitude $40\deg$, as a proxy of the actual angular velocity,
in place of $\Delta f$, resulting in an effective Rossby number $\Ro_{eff}$. 
In order to mimic a weaker electric coupling, numerical simulations have been run where the electric conductivity
of the inner sphere is reduced as compared to that of copper (nominal conductivity ratio of 4.2). Synthetic angular velocity profiles are shown in
figure \ref{fig:synthetics}a for a conductivity ratio of 0.04 and 0.01. The profiles retain the same shape but the overall
amplitude is severely reduced. This explains why we could get coherent results by renormalizing all data with the
$\Delta V_{40\deg}$ proxy. Comparing the spread in figure \ref{fig:ddp40_vs_df} with the numerical simulations suggests
that the effective conductivity at the sodium/copper interface could be reduced by as much as 100 by wetting
problems. We observe that the electric coupling deteriorates with time over the months, but that it can be somewhat revived
by letting the copper in contact with liquid sodium at $150 \deg$C during 3 days.

The numerical simulations show that the meridional circulation is not very sensitive to the conductivity
ratio. Radial profiles of the radial velocity are shown in figure \ref{fig:synthetics}b for the numerical
model with the conductivity ratio of 0.04. As in figure \ref{fig:numerics}b, the circulation is centrifugal
at the equator ($\Ro>0$) and the radial velocity is positive at the latitude of $10\deg$. It is negative
at $-40\deg$, in good agreement with our measurements (figure \ref{fig:Ur}b). At $20\deg$, the profile is
more complex, but it is also negative near the inner sphere, as experimentally observed (figure \ref{fig:Ur}a).
In the numerical model, the radial velocities reach about $1\%$ of the inner sphere tangential velocity, while it is about twice as large in the experiments. The experiments show clear evidence
of a secondary meridional cell farther away from the inner sphere.

\subsection{The peculiar regime for $\Ro_{eff} \simeq -1$}
\label{Pic_Nadege}

In the counter--rotating case, the centripetal circulation at the equator, which characterizes the moderate Rossby
number regime, transforms into a centrifugal circulation at the equator as the Rossby number increases. We recall that at small
Rossby number, the meridional circulation is governed by Hartmann--pumping at the inner sphere, and is therefore
directed towards the inner sphere when it is counter--rotating, while centrifugation from the inner sphere takes over at large Rossby numbers. The transition
between the two regimes occurs over a narrow interval of $\Delta V_{40\deg}$, and is marked by a peak of the
induced magnetic field. This happens where the effective angular velocity of the fluid is such that it almost cancels that
of the outer sphere, as seen in the laboratory frame.

Our interpretation is that, in a narrow window of $\Delta V_{40\deg}$, the antagonistic effects of the rotation of the outer
and inner spheres drive the liquid in a state that shows almost no rotation with respect to the laboratory frame. The constraint of rotation then
vanishes, and a vigorous meridional circulation is allowed, resulting in an enhanced induced magnetic field.

\begin{figure}
  \centerline{ \includegraphics[width=10cm]{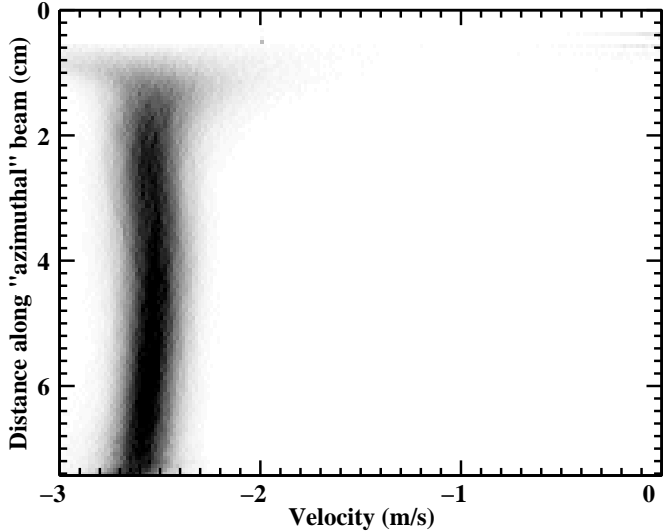}}

     \caption{A profile ($pdf$) of the angular velocity of the fluid for $f = 4.2$Hz and $\Delta f = -18.2$Hz, for which $\Delta V_{40\deg} = -1.7$mV. This value
	 corresponds to the peak observed for the induced magnetic field.}
        \label{fig:uphi_peak_B}
\end{figure}

Some support for this interpretation is given by a direct measurement of the angular velocity using the ultrasonic Doppler technique. As shown
in figure \ref{fig:uphi_peak_B}, a peculiar profile is measured close to the $B$ peak. An almost uniform angular velocity is measured (down to $d \simeq 7.4$ cm, i.e. at mid--depth of the fluid shell). Using equation \ref{eq:umes}, we find a fluid rotation frequency of -5.7 Hz on this plateau. Since the
ultrasonic probe is installed in the rotating sphere, the angular velocity with respect to the laboratory frame is obtained by adding the
angular velocity of the outer sphere. The resulting angular velocity is then -5.7 + 4.2 = -1.5 Hz, which is small but non zero.
Note that the electric potentials (figure \ref{fig:ddp_vs_ddp40}) also suggest that, on and beyond
the peak, the fluid is nearly in solid--body rotation.
However, the above scenario is probably too simple, and the variation of the induced magnetic field calls for a more sophisticated
interpretation. For example, the $B_r$ and $B_{\theta}$ records of figure \ref{fig:run_type} show that the ``centrifugal'' regime
is announced by a strong rise of $B_{\theta}$, while $B_r$ starts decreasing instead. Besides, pure solid--body
rotation would imply no meridional circulation altogether.

\subsection{Implications for experimental dynamos}
The transition between the centripetal and the centrifugal regimes deserves a specific discussion
in the context of the preparation of experimental dynamos. In \cite{Cardin02}, we evaluated the possibility to run a dynamo
where the forcing is a rotating spherical Couette flow. Such a dynamo would run in the magnetostrophic regime, expected for
planetary cores, and it would be of the $\alpha \omega$--type, displaying large azimuthal magnetic fields. However, the critical
magnetic Reynolds number $\Rm_c$ for these dynamos appears to be high \citep{Schaeffer06}.
Indeed, there is an optimum ratio of meridional to azimuthal velocities that yield a minimum $\Rm_c$ \citep{Ravelet05,Bayliss07}. Here, we find that, in
a rapidly rotating spherical Couette flow,
the meridional velocities are at least one order of magnitude smaller than the azimuthal velocities.
In contrast,
the mean flow in other experiments \citep{Peffley00,Ravelet05,Bayliss07} is tuned to be close to an axi--symmetric flow of the \cite{Dudley89} family, with
a strong meridional circulation, yielding an $\alpha^2$--dynamo, and moderate $\Rm_c$ \citep{Kumar75,Dudley89}.
It has been recently suggested that the dynamo efficiency of such flows can be
significantly reduced by turbulent fluctuations \citep{Bayliss07}.
Now, in the narrow
window that corresponds to the transition from the centripetal regime to the centrifugal regime, the magnitude of the
meridional circulation
becomes comparable to the azimuthal one. Thus, the value of the magnetic Reynolds number for the onset of dynamo action may be minimum around this transition. A rotating spherical Couette flow dynamo experiment should
thus make it possible to observe the transition from an $\alpha^2$--type (with $\Rm_c$
perhaps large compared to its laminar value) to an $\alpha \omega$--type dynamo
(with a large $\Rm_c$ for the laminar flow but weak turbulent effects). We are currently
conducting extensive numerical simulations of the dynamo instability for spherical Couette flow to document this effect.

\subsection{Implications for core dynamics}
The modified Taylor state observed in the $DTS$ experiment bears some relevance to the expected dynamical regime in the
Earth's core. Indeed, the Ekman number based on the expected fluid viscosity is very small, but observations of the nutations
reveal that the friction at the core--mantle boundary is some $10^4$ larger than expected for linear Ekman friction
\citep{Herring02,Koot08}. Although the origin of this enhanced friction is still debated \citep{Deleplace06,Buffett07} and is
undoubtedly different from the turbulent friction present in $DTS$, it would
be worth investigating its effect on the dynamics of the core, from a modified Taylor state point of view. Quasi--geostrophic
numerical models \citep{Schaeffer06}, in which Ekman friction is parameterized, are a particularly attractive tool for
quantifying this effect. 

\section*{\bf Acknowledgments}
We thank Jean-Paul Masson and Patrick La Rizza for their skillful technical assistance, and Alexandre Fournier
for useful discussions. We are indebted to two anonymous reviewers whose comments helped us improving the manuscript. The $DTS$ project is supported by Fonds National de la Science, Institut National des Sciences de l'Univers, Centre National de la Recherche Scientifique, R\'egion Rh\^ one-Alpes and Universit\'e Joseph Fourier. This research benefited from an invitation of HCN and DJ at the KITP, Santa Barbara, and was thus supported in part by the USA National Science Foundation under Grant No. PHY05-51164.


\begin{thebibliography}{34}
\expandafter\ifx\csname natexlab\endcsname\relax\def\natexlab#1{#1}\fi
\expandafter\ifx\csname url\endcsname\relax
  \def\url#1{\texttt{#1}}\fi
\expandafter\ifx\csname urlprefix\endcsname\relax\def\urlprefix{URL }\fi

\bibitem[{{Bayliss} et~al.(2007){Bayliss}, {Forest}, {Nornberg}, {Spence}, and
  {Terry}}]{Bayliss07}
{Bayliss}, R.~A., {Forest}, C.~B., {Nornberg}, M.~D., {Spence}, E.~J., {Terry},
  P.~W., Feb. 2007. {Numerical simulations of current generation and dynamo
  excitation in a mechanically forced turbulent flow}. Physical Review E
  75~(2), 026303--+.

\bibitem[{{Braginsky} and {Meytlis}(1990)}]{Braginsky90}
{Braginsky}, S.~I., {Meytlis}, V.~P., 1990. {Local turbulence in the Earth's
  core}. Geophysical and Astrophysical Fluid Dynamics 55, 71--87.

\bibitem[{{Brito} et~al.(2001){Brito}, {Nataf}, {Cardin}, {Aubert}, and
  {Masson}}]{Brito01}
{Brito}, D., {Nataf}, H.-C., {Cardin}, P., {Aubert}, J., {Masson}, J.-P., 2001.
  {Ultrasonic Doppler velocimetry in liquid gallium}. Experiments in Fluids 31,
  653--663.

\bibitem[{{Buffett} and {Christensen}(2007)}]{Buffett07}
{Buffett}, B., {Christensen}, U., 2007. {Magnetic and viscous coupling at the
  core--mantle boundary: inferences from observations of the Earth's
  nutations}. Geophys. J. Int. 171, 145--152.

\bibitem[{{Buffett}(2003)}]{Buffett03}
{Buffett}, B.~A., Jun. 2003. {A comparison of subgrid-scale models for
  large-eddy simulations of convection in the Earth's core}. Geophysical
  Journal International 153, 753--765.

\bibitem[{{Cardin} et~al.(2002){Cardin}, {Brito}, {Jault}, {Nataf}, and
  {Masson}}]{Cardin02}
{Cardin}, P., {Brito}, D., {Jault}, D., {Nataf}, H.-C., {Masson}, J.-P., 2002.
  {Towards A Rapidly Rotating Liquid Sodium Dynamo Experiment}.
  Magnetohydrodynamics 38, 177--189.

\bibitem[{{Deleplace} and {Cardin}(2006)}]{Deleplace06}
{Deleplace}, B., {Cardin}, P., 2006. {Viscomagnetic torque at the core--mantle
  boundary}. Geophys. J. Int. 167, 557--566.

\bibitem[{{Dormy} et~al.(1998){Dormy}, {Cardin}, and {Jault}}]{Dormy98}
{Dormy}, E., {Cardin}, P., {Jault}, D., Jul. 1998. {MHD flow in a slightly
  differentially rotating spherical shell, with conducting inner core, in a
  dipolar magnetic field}. Earth and Planetary Science Letters 160, 15--30.

\bibitem[{{Dormy} et~al.(2002){Dormy}, {Jault}, and {Soward}}]{Dormy02}
{Dormy}, E., {Jault}, D., {Soward}, A.~M., 2002. {A super--rotating shear layer
  in magnetohydrodynamic spherical Couette flow}. J. Fluid Mech. 452, 263--291.

\bibitem[{{Dudley} and {James}(1989)}]{Dudley89}
{Dudley}, M.~L., {James}, R.~W., Oct. 1989. {Time-dependent kinematic dynamos
  with stationary flows}. Royal Society of London Proceedings Series A 425,
  407--429.

\bibitem[{{Eckert} and {Gerbeth}(2002)}]{Eckert02}
{Eckert}, S., {Gerbeth}, G., 2002. {Velocity measurements in liquid sodium by
  means of ultrasound Doppler velocimetry}. Experiments in Fluids 32, 542--546.

\bibitem[{{Fearn} et~al.(1988){Fearn}, {Roberts}, and {Soward}}]{Fearn88}
{Fearn}, D.~R., {Roberts}, P.~H., {Soward}, A.~M., 1988. {Convection, stability
  and the dynamo}. In: {Galdi}, G.~P., {Straughan}, B. (Eds.), Energy Stability
  and Convection. Longman scientific and technical Harlow, pp. 60--324.

\bibitem[{{Ferraro}(1937)}]{Ferraro37}
{Ferraro}, V.~C.~A., 1937. {The non-uniform rotation of the sun and its
  magnetic field}. Month. Notices Roy. Astr. Soc. 97, 458--472.

\bibitem[{{Herring} et~al.(2002){Herring}, {Mathews}, and
  {Buffett}}]{Herring02}
{Herring}, T., {Mathews}, P., {Buffett}, B., 2002. {Modeling nutation and
  precession: very long baseline interferometry results}. J. Geophys. Res.
  107B, 145--152.

\bibitem[{{Hollerbach}(1994)}]{Hollerbach94}
{Hollerbach}, R., 1994. {Magnetohydrodynamic Ekman and Stewartson layers in a
  rotating spherical shell}. Proc. R. Soc. Lond. A 444, 333--346.

\bibitem[{{Hollerbach}(2000)}]{Hollerbach00}
{Hollerbach}, R., 2000. {Magnetohydrodynamic flows in spherical shells}. In:
  {Egbers}, C., {Pfister}, G. (Eds.), LNP Vol. 549: Physics of Rotating Fluids.
  pp. 295--+.

\bibitem[{{Hollerbach} et~al.(2007){Hollerbach}, {Canet}, and
  {Fournier}}]{Hollerbach07}
{Hollerbach}, R., {Canet}, E., {Fournier}, A., Feb. 2007. {Spherical Couette
  flow in a dipolar magnetic field}. European J. Mech. Fluids 26, 729--737.

\bibitem[{{Jault}(2008)}]{Jault08}
{Jault}, D., 2008. {Axial invariance of rapidly varying diffusionless motions
  in the Earth's core interior}. Physics of the Earth and Planetary Interiors
  166, 67--76.

\bibitem[{{Kelley} et~al.(2007){Kelley}, {Triana}, {Zimmerman}, {Tilgner}, and
  {Lathrop}}]{Kelley07}
{Kelley}, D.~H., {Triana}, S.~A., {Zimmerman}, D.~S., {Tilgner}, A., {Lathrop},
  D.~P., 2007. {Inertial waves driven by differential rotation in a planetary
  geometry}. Geophysical and Astrophysical Fluid Dynamics 101, 469--487.

\bibitem[{{Kleeorin} et~al.(1997){Kleeorin}, {Rogachevskii}, {Ruzmaikin},
  {Soward}, and {Starchenko}}]{Kleeorin97}
{Kleeorin}, N., {Rogachevskii}, I., {Ruzmaikin}, A., {Soward}, A.~M.,
  {Starchenko}, S.~V., 1997. {Axisymmetric flow between differentially rotating
  spheres in a dipole field}. J. Fluid Mech. 344, 213--244.

\bibitem[{{Koot} et~al.(2008){Koot}, {Rivoldini}, {de Viron}, and
  {Dehant}}]{Koot08}
{Koot}, L., {Rivoldini}, A., {de Viron}, O., {Dehant}, V., 2008. {Estimation of
  Earth interior parameters from a Bayesian inversion of VLBI nutation time
  series}. Journal of Geophysical Research, in press.

\bibitem[{{Kumar} and {Roberts}(1975)}]{Kumar75}
{Kumar}, S., {Roberts}, P.~H., 1975. {A Three-Dimensional Kinematic Dynamo}.
  Royal Society of London Proceedings Series A 344, 235--258.

\bibitem[{{Lehnert}(1954)}]{Lehnert54}
{Lehnert}, B., May 1954. {Magnetohydrodynamic Waves Under the Action of the
  Coriolis Force}. Astrophysical Journal 119, 647--+.

\bibitem[{{Matsui} and {Buffett}(2007)}]{Matsui07}
{Matsui}, H., {Buffett}, B.~A., 2007. {Commutation error correction for large
  eddy simulations of convection driven dynamos}. Geophysical and Astrophysical
  Fluid Dynamics 101, 429--449.

\bibitem[{{Nataf} et~al.(2006){Nataf}, {Alboussi{\`e}re}, {Brito}, {Cardin},
  {Gagni{\`e}re}, {Jault}, {Masson}, and {Schmitt}}]{Nataf06}
{Nataf}, H.-C., {Alboussi{\`e}re}, T., {Brito}, D., {Cardin}, P.,
  {Gagni{\`e}re}, N., {Jault}, D., {Masson}, J.-P., {Schmitt}, D., Oct. 2006.
  {Experimental study of super-rotation in a magnetostrophic spherical Couette
  flow}. Geophysical and Astrophysical Fluid Dynamics 100, 281--298.

\bibitem[{{Peffley} et~al.(2000){Peffley}, {Goumilevski}, {Cawthrone}, and
  {Lathrop}}]{Peffley00}
{Peffley}, N.~L., {Goumilevski}, A.~G., {Cawthrone}, A.~B., {Lathrop}, D.~P.,
  Jul. 2000. {Characterization of experimental dynamos}. Geophysical Journal
  International 142, 52--58.

\bibitem[{{Ravelet} et~al.(2005){Ravelet}, {Chiffaudel}, {Daviaud}, and
  {L{\'e}orat}}]{Ravelet05}
{Ravelet}, F., {Chiffaudel}, A., {Daviaud}, F., {L{\'e}orat}, J., Nov. 2005.
  {Toward an experimental von K{\'a}rm{\'a}n dynamo: Numerical studies for an
  optimized design}. Physics of Fluids 17, 7104--+.

\bibitem[{{Schaeffer} and {Cardin}(2006)}]{Schaeffer06}
{Schaeffer}, N., {Cardin}, P., May 2006. {Quasi-geostrophic kinematic dynamos
  at low magnetic Prandtl number}. Earth and Planetary Science Letters 245,
  595--604.

\bibitem[{{Schmitt} et~al.(2008){Schmitt}, {Alboussi{\`e}re}, {Brito},
  {Cardin}, {Gagni{\`e}re}, {Jault}, and {Nataf}}]{Schmitt08}
{Schmitt}, D., {Alboussi{\`e}re}, T., {Brito}, D., {Cardin}, P.,
  {Gagni{\`e}re}, N., {Jault}, D., {Nataf}, H.-C., 2008. {Rotating spherical
  Couette flow in a dipolar magnetic field: Experimental study of
  magneto--inertial waves}. Journal of Fluid Mechanics 604, 175--197.

\bibitem[{{Sisan} et~al.(2004){Sisan}, {Mujica}, {Tillotson}, {Huang},
  {Dorland}, {Hassam}, {Antonsen}, and {Lathrop}}]{Sisan04}
{Sisan}, D.~R., {Mujica}, N., {Tillotson}, W.~A., {Huang}, Y.-M., {Dorland},
  W., {Hassam}, A.~B., {Antonsen}, T.~M., {Lathrop}, D.~P., Sep. 2004.
  {Experimental Observation and Characterization of the Magnetorotational
  Instability}. Physical Review Letters 93~(11), 114502--+.

\bibitem[{{St.~Pierre}(1996)}]{StPierre96}
{St.~Pierre}, M.~G., 1996. {On the local nature of turbulence in Earth's outer
  core}. Geophysical and Astrophysical Fluid Dynamics 83, 293--306.

\bibitem[{{Starchenko}(1998)}]{Starchenko98}
{Starchenko}, S.~V., 1998. {Magnetohydrodynamic flow between insulating shells
  rotating in strong potential field}. Physics of Fluids 10, 2412--2420.

\bibitem[{{Stewartson}(1966)}]{Stewartson66}
{Stewartson}, K., 1966. {On almost rigid rotations. Part 2.} Journal of Fluid
  Mechanics 26, 131--144.

\bibitem[{{Taylor}(1963)}]{Taylor63}
{Taylor}, J., 1963. {The magneto--hydrodynamics of a rotating fluid and the
  Earth's dynamo problem}. Proc. R. Soc. Lond. A 274, 274--283.

\end{thebibliography}

\end{document}